\documentclass[a4paper]{ar-1col}
\usepackage{arydshln}
\usepackage[numbers]{natbib}
\usepackage{bm}
\usepackage{url}
\usepackage{amsmath,amssymb}
\usepackage{pifont}
\usepackage{subfigure}
\usepackage[dvipsnames]{xcolor}
\usepackage{multirow}

\usepackage{amssymb}

\usepackage{tikz}
\usetikzlibrary{decorations.markings}

\newcommand{\bea}{\begin{eqnarray}}
\newcommand{\eea}{\end{eqnarray}}

\def\gs{\text{gs}}

\def\up{\uparrow}
\def\down{\downarrow}

\def\CO{{\cal O}}
\def\bra#1{\left\langle#1\right|}
\def\ket#1{\left|#1\right\rangle}

\def\vev#1{\left\langle{#1}\right\rangle}

\def\apd{{\vphantom{\dagger}}}

\def\thump{\hfill$\blacksquare$}

\def\parfig#1#2{
\parbox{#1\textwidth}
{\includegraphics[width=#1\textwidth]{#2}}
}

\def\dd{\text{d}}
\def\dbar{{\mathchar'26\mkern-12mu \dd}} 

\def\CO{{\cal O}}
\def\IZ{\mathbb{Z}}
\def\IR{\mathbb{R}}

\def\CA{{\cal A}}
\def\CB{{\cal B}}
\def\CL{{\cal L}}
\def\CN{{\cal N}}

\def\ii{{\bf i}}

\def\({\left(}
\def\){\right)}
\def\be{\begin{equation}}
\def\ee{\end{equation}}

\def\nd{{\vphantom{\dagger}}}

\def\bra#1{\left\langle#1\right|}
\def\ket#1{\left|#1\right\rangle}

\def\vev#1{\left\langle{#1}\right\rangle}
\def\ketbra#1#2{ | #1 \rangle\hskip-2pt\langle #2|}

\def\gSO{\textsf{SO}}

\def\gSU{\textsf{SU}}
\def\gU{\textsf{U}}

\def\gs{\text{gs}}

\def\tr{{\rm tr}}

\def\eqref#1{Eq.~\ref{#1}}

\usepackage[
      linkcolor=Mathematica1,
      colorlinks=true,
      urlcolor=Mathematica2,
      filecolor=black,
      citecolor=Mathematica3,
      pdftitle={},
        pdfauthor={John McGreevy},
        pdfsubject={},
        pdfkeywords={},
      ]{hyperref}

\def\ie{{\it i.e.\/}}

\def\eg{{\it e.g.\/}}
\def\frac#1#2{{\textstyle{#1 \over #2}}}
\def\half{\frac{1}{2}}

\def\nd{^{\vphantom{\dagger}}}

\def\eps{\epsilon}

\def\gtwid{\,{\raise.3ex\hbox{$>$\kern-.75em\lower1ex\hbox{$\sim$}}}\,}

\def\CO{{\cal O}}

\def\det{\textsf{det}}

\def\ket#1{{\big| \, #1\, \big\rangle}}
\def\bra#1{{ \big\langle \, #1 \, \big|}}

\def\HH{{\hat H}}

\definecolor{Mathematica1}{rgb}{0.368417, 0.506779, 0.709798}
\definecolor{Mathematica2}{rgb}{0.880722, 0.611041, 0.142051}
\definecolor{Mathematica3}{rgb}{0.560181, 0.691569, 0.194885}

\definecolor{JMblue}{RGB}{25,25,125}

\definecolor{darkblue}{rgb}{0,0,0.4}

\jname{Submitted to Annual Review of Condensed Matter Physics}
\jyear{2022}

\begin{document} 

\markboth{McGreevy}{Generalized Symmetries}

\title{Generalized Symmetries in Condensed Matter}
\author{John McGreevy
\affil{Department of Physics, University of California San Diego, La Jolla, California 92093, USA}
}

\begin{abstract}
Recent advances in our understanding of symmetry in quantum many-body systems 
offer the possibility of a generalized Landau paradigm that encompasses all equilibrium phases of matter.  
This is a brief and elementary review of some of these developments.
\end{abstract}

\maketitle

\tableofcontents

\section{Extending the Landau Paradigm}

If you have been to a condensed-matter talk in the past few decades, you have seen the beating that Landau has been taking.   
The speaker begins by saying that Landau told us that states of matter are classified by the symmetries they break.
After showing a picture of a donut, the speaker explains that in {\it this} talk, in contrast, they will discuss a state of matter that goes beyond Landau's limited conception of the world.  

Having given such talks myself, I think it is extremely interesting that, in fact, with modern generalizations of our understanding of symmetry, it may be possible to incorporate all known equilibrium phases of matter into a suitably extended version of the Landau Paradigm.
Let me attempt to paraphrase the Landau Paradigm:
\begin{enumerate}
\item \label{item1} Phases of matter should be labelled by how they represent their symmetries, in particular whether they are spontaneously broken or not.

\item \label{item2} The degrees of freedom at a critical point are the fluctuations of the order parameter.

\end{enumerate}
A significant corollary of assertion \ref{item1} is that gapless degrees of freedom, or groundstate degeneracy, in a phase,
should be swept out by a symmetry.  That is, they should arise as Goldstone modes for some spontaneously broken symmetry.

 Beyond its conceptual utility, this perspective
 has a weaponization, in the form of Landau-Ginzburg theory, in terms of which 
we may find representative states, understand gross phase structure, and, when suitably 
augmented by the renormalization group (RG), even quantitatively describe phase transitions.

Indeed there are many apparent exceptions to the Landau Paradigm.  Let us focus first on apparent exceptions to item \ref{item1}.
As a preview, exceptions that are only apparent include:
\begin{itemize}
\item {\bf Topologically-ordered states.} These are phases of matter distinguished from the trivial phase by something other than a local order parameter \cite{WEN1990a,Wegner1971}.
Symptoms include a groundstate degeneracy that depends on the topology of space, and {\it anyons}, excitations that cannot be created by any local operator.  
Real examples found so far include fractional quantum Hall states, as well as gapped spin liquids.

\item {\bf Other deconfined states of gauge theory.} This category includes gapless spin liquids 
such as spinon Fermi surface or Dirac spin liquids (most candidate spin liquid materials are gapless).  
Another very visible manifestation of such a state is the photon phase of quantum electrodynamics in which our vacuum lives.

\item {\bf Fracton phases.} Gapped fracton phases are a special case of topological order, where there are excitations that not only cannot be {\it created} by any local operator, but  cannot be {\it moved} by any local operator.

\item {\bf Topological insulators.}  Here we can include both free-fermion states with topologically non-trivial bandstructure, as well as interacting symmetry-protected topological (SPT) phases.

\item {\bf Landau Fermi liquid.}

\end{itemize}

{\bf Conventions.}  $L$ is the linear system size.  
$D=d+1$ is the number of spacetime dimensions.
I'll denote the dimension of a manifold or the degree of a form by a subscript or superscript.
I will use fancy upper-case letters (like ${\cal A}_\mu$) for background gauge fields
and lower-case letters (like $a_\mu$) for dynamical gauge fields.
I will sometimes use $G^{(p)}$ to denote a $p$-form symmetry with group $G$.

\begin{figure}[h!]
$$\parfig{.45}{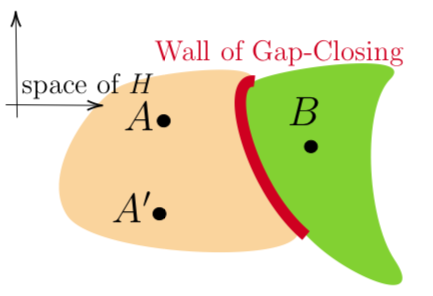}\parfig{.45}{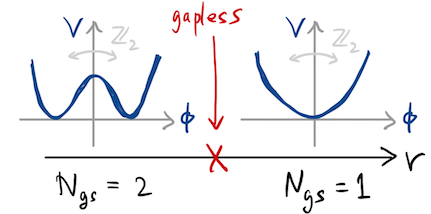}$$
\caption{Left: A schematic illustration of the definition of gapped phases of matter.  Two distinct phases are separated in the space of local Hamiltonians by a wall of gaplessness, the codimension-one locus where the gap closes.  Here $H_{A} \simeq H_{A'}$.  
Right: The groundstate degeneracy, $N_\text{gs}$, for example as swept out by a spontaneously broken global symmetry, is an example of a topological invariant that can label a phase.}
\label{fig:gapped-phases}
\end{figure}

{\bf Brief non-symmetry-based accounting of gapped phases.}
A useful definition of a gapped phase of matter 
is an equivalence class of gapped groundstates of local Hamiltonians, in the thermodynamic limit\footnote{We should pause to comment on the meaning of `gapped'.  
We allow for a stable groundstate subspace, which becomes degenerate in the thermodynamic limit.  
`Stable' means that the degeneracy persists under arbitrary small perturbations of the Hamiltonian, and requires that the groundstates are not related by the action of local operators.
In $d$ spatial dimensions, the logarithm of the number of such states can grow as quickly as $L^{d-1}$ \cite{Haah:2020ghp} in fracton models.
}.
Two groundstates are considered equivalent if they are related by adiabatic evolution (for a time of order $L^0$) combined with inclusion or removal of product states.  
That is, there is a path between the two Hamiltonians along which the gap does not close (see Fig.~\ref{fig:gapped-phases}, left).

This definition poses a difficulty for checking that two Hamiltonians represent distinct phases: we cannot check all possible paths between them.  A crucial role is therefore played by universal properties of a phase -- quantities, such as integers, that cannot change smoothly within a phase, and therefore can only vary across phase boundaries.  A good example of such a topological invariant is the groundstate degeneracy, which is certainly an integer.
A phase of matter that spontaneously breaks a discrete symmetry $G$ has a groundstate degeneracy $|G|$, the order of the group (see Fig.~\ref{fig:gapped-phases}, right).  
This is a topological distinction from the trivial paramagnetic phase, which has a unique groundstate and a representative that is a product state with no entanglement at all.  In this sense, even spontaneous symmetry breaking (SSB) is a topological phenomenon.

Non-trivial phases can be divided into two classes: those with topological order and those without.  
One way to define topological order \cite{WEN1990a} is a phase with localized excitations that cannot be created by any local operator.  
In 2+1 dimensions, such particle excitations are called {\it anyons}; they can be created in pairs by an open-string operator.  
On a space with a non-contractible curve $C$, new groundstates can be made by acting with the operator that transports an anyon around $C$.
These groundstates are {\it locally indistinguishable}, in the following sense.
If $\ket{\parfig{.05}{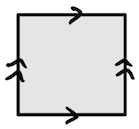}}$ and $\ket{\parfig{.05}{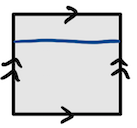}}$ are 
two such groundstates, then
\be \label{eq:indistinguishable} \bra{\parfig{.05}{figs/fig-TC-torus10.png}} \CO_x \ket{\parfig{.05}{figs/fig-TC-torus00.png}} = 0 ,\ee 
for all local operators $\CO_x$.
(The picture in the kets is a cartoon of two of the groundstates on the 2-torus.)
A final symptom is the existence of long-range entanglement in the groundstate; a review focussing on this aspect is 
\cite{grover2013entanglement}.

An interesting special case of topologically ordered states is fracton phases \cite{nandkishore2019fractons,pretko2020fracton}.  A fracton phase has excitations that cannot be {\it moved} by any local operator (perhaps only in some directions of space).  This is a strictly stronger condition than topological order, since an excitation can effectively be moved by annihilating it and creating it again elsewhere.  
Such phases (with a gap) exist in 3+1 dimensions (and higher).  A consequence of the defining property is a groundstate degeneracy
whose logarithm grows linearly with system size, and a subleading linear term in the scaling of the entanglement entropy of a region with the size of the region.

Even without topological order, there can be phases distinct from the trivial phase.  
One way in which they can be distinguished is by what happens if we put them on a space with boundary, so that there is a spatial interface with the trivial phase.
A very rough (and not entirely correct) idea is that if the gap must close along the path to the trivial phase, then the coupling must pass through the wall of gap-closing at the edge of the sample.  Phases that are distinguished in this way include integer quantum Hall states, topological insulators, and, more generally,
symmetry-protected topological (SPT) phases such as the Haldane phase of the spin-one chain, or polyacetylene.

It seems that all of these examples transcend the Landau Paradigm.  
My goal here is not to use the Landau Paradigm as a straw man, but rather to pursue it in earnest.
The idea is that by suitably refining and generalizing our notions of symmetry, we can incorporate all of these `beyond-Landau' examples into a {\it Generalized Landau Paradigm}.  
There are two crucial ingredients, which work in concert: anomalies and generalized symmetries.

In this article, I am speaking of actual symmetries of physical systems, sometimes called `global symmetries'.  
They act on the Hilbert space and take one state to another.
In contrast, there is no such thing as `gauge symmetry'.
In a gauge theory, the gauge invariance is a redundancy of a particular description of the system, 
and is not preserved by relabelling degrees of freedom.
For example, dualities (equivalences of physical observables at low energies) often relate a gauge theory with one gauge group 
to a gauge theory with a distinct gauge group.
A familiar example in condensed matter physics is the duality between the XY model and the abelian Higgs model 
in 2+1 dimensions \cite{Peskin:1977kp,Dasgupta:1981zz}, but there are many others, \eg~\cite{Seiberg:1994pq}.
This complaint about terminology hides an abyss of human ignorance:
If someone hands you a piece of rock and asks whether its low-energy physics is described by some phase of a gauge theory, 
how will you tell?
It is certainly true that phases realizable by gauge theory go beyond other constructions with only short-ranged entanglement;
this begs for a characterization of these phases that transcends a description in terms of redundancies.  
Higher-form symmetries offer such a characterization for some such phases.

I want to highlight early attempts to understand topological order
\cite{Nussinov:2006iva, Nussinov:2009zz},
and the gaplessness of the photon
\cite{Kovner:1992pu}
as a consequences of generalized symmetry, as well as 
early appearances of generalized symmetries in the string theory literature
\cite{Pantev:2005rh, Hellerman:2006zs, Sharpe:2015mja}.
Other papers that have explicitly advocated for the utility of a generalized Landau Paradigm 
include \cite{Gaiotto:2014kfa, Hofman:2018lfz,Delacretaz:2019brr, Iqbal:2020msy,2021PhRvR...3c3024Z}.

\section{Higher-form symmetries}

The concept of higher-form symmetry that we review here was explained in \cite{Kapustin:2013uxa, Gaiotto:2014kfa}.
It is easiest to introduce using a relativistic notation.  Indices $\mu, \nu$ run over space and time.

Let's begin by considering the familiar case of a continuous 0-form symmetry.
Noether's theorem guarantees a conserved current $J_\mu$ satisfying
 $ \partial^\mu J_\mu  = 0 $.
 In the useful language of differential forms, this is $d \star J = 0 $, where $\star$ is the Hodge duality operation\footnote{The Hodge dual of a $p$-form $\omega_p$ on a $D$-dimensional space with metric $g_{\mu\nu}$ has components $(\star \omega_p)_{\mu_{1} \cdots \mu_{D-p} } 
 = \sqrt{\det g} \epsilon_{\mu_1 \cdots \mu_D} \omega_p^{\mu_{D-p+1} \cdots  \mu_D}$, where indices are raised with the inverse metric $g^{\mu\nu}$ and $\epsilon_{\mu_1 \cdots \mu_D}$ is the antisymmetric Levi-Civita symbol.}.
This continuity equation has the consequence that the {\it charge}
 $ Q_\Sigma = \int_{\Sigma_{D-1} } \star J $ is independent of the choice of time-slice $\Sigma$.
 ($\Sigma$ here is a closed $d$-dimensional surface, of codimension one in spacetime.)
 Notice that this is a topological condition.
$Q_\Sigma$ commutes with the Hamiltonian, 
 the generator of time translations, 
 and therefore so does the unitary operator $ U_\alpha = e^{ i \alpha Q}$,
 which we call the {\it symmetry operator}\footnote{Throughout I will assume that the normalization is such that $ Q \in  \IZ$, so
  that $\alpha \equiv \alpha + 2\pi$.}.

\begin{figure}[h!]
$$\parfig{.45}{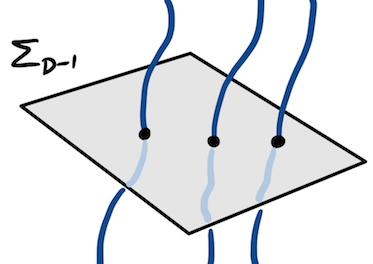}~~~\parfig{.37}{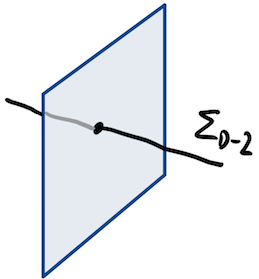}$$
\caption{Left: In the case of an ordinary 0-form symmetry, the charge is integrated over a codimension-one slice of spacetime $\Sigma_{D-1}$, often a slice of constant time.  All the particle worldlines (blue curves) must pass through this hypersurface.
Right: The charge of a 1-form symmetry is integrated over a codimension-two locus of spacetime $\Sigma_{D-2}$ (a string in the case of $D=2+1$).  
This surface intersects the worldsheets of strings (blue sheet).
}
\label{fig:sigmas}
\end{figure}

If the charge is carried by particles, $Q_\Sigma$ counts the number of particle worldlines piercing the surface $\Sigma$ (as in Fig.~\ref{fig:sigmas}, left), and the conservation law  $\dot Q=0$ says that 
charged particle worldlines cannot end
except on charged operators.
If instead of a $\gU(1)$ symmetry,
we only had a discrete $\IZ_p$ symmetry 
we could simply restrict $\alpha \in \{0, 2\pi/p, 4 \pi/p ... (p-1) 2\pi/p\}$ in the symmetry operator $U_\alpha$.
In that case, particles can disappear in groups of $p$.

Objects charged under a 0-form symmetry are created by local operators.
Local operators transform under the symmetry by 
$ \CO(x) \to U_\alpha \CO(x) U_\alpha^\dagger = e^{ \ii q \alpha} \CO(x), d\alpha=0 $, where $q$ is the charge of the operator.
The infinitesimal version is:
$ \delta \CO(x) = \ii [ Q, \CO(x)] = \ii q \CO(x)$.

Now let us consider a continuous 1-form symmetry.  This means that there 
is a conserved current which has two indices, and is completely antisymmetric: 
\be \label{eq:1-form-conservation} J_{\mu\nu} = - J_{\nu\mu}~\text{with}~\partial^\mu J_{\mu\nu} = 0 .\ee
We can regard $J$ as a 2-form and write the conservation law \eqref{eq:1-form-conservation} as $d \star J = 0 $.
As a consequence, for any closed codimension-{\it two} locus in spacetime $\Sigma_{D-2}$, the quantity 
$Q_\Sigma = \int_{\Sigma_{D-2}} \star J $ depends only on the topological class of $\Sigma$. 
The analog of the symmetry operator is the unitary operator 
\be \label{eq:1-form-symmetry-op} U_\alpha(\Sigma) = e^{ \ii \alpha Q_\Sigma} .\ee
Notice that reversing the orientation of $\Sigma$ produces the adjoint of $U$:
$ U_\alpha(-\Sigma) =  U_\alpha^\dagger(\Sigma)$.

The charge $Q_\Sigma$ in the 1-form case counts the number of charged string worldsheets intersecting the surface $\Sigma$ (as in Fig.~\ref{fig:sigmas}, right).
The conservation law \eqref{eq:1-form-conservation} then says that 
charged string worldsheets cannot end except on charged operators. 
The objects charged under a 1-form symmetry are loop operators, $W(C)$.  
Fixing a constant-time slice $M_{D-1}$, such a loop operator transforms as 
\be W(C)\to 
U_\alpha(\Sigma) W(C) U_\alpha^\dagger(\Sigma) = 
e^{ \ii \alpha \oint_C \Gamma_\Sigma } W(C), ~~~~d \Gamma_\Sigma =0 .\ee
Here $\Sigma_{D-2}\subset M_{D-1} $ is any closed $(D-2)$-manifold, and $\Gamma_\Sigma$ is its Poincar\'e dual in $M_{D-1}$,
in the sense that $\int_{M_{D-1}} \eta^{(D-2)} \wedge \Gamma_\Sigma = \int_{\Sigma_{D-2}} \eta^{(D-2)} $ 
for all $\eta$; $d \Gamma_\Sigma=0$ because $\Sigma$ has no boundary.  
The infinitesimal version of this transformation law is 
\be \delta W(C) = \ii [ Q_\Sigma, W(C)] = \ii q \#\( \Sigma,C\) W(C) ,\ee
where $\#(\Sigma, C)$ is the intersection number in $M$\footnote{Above I have written the expression for the transformation as $U(\Sigma) W(C) U^\dagger(\Sigma)$.  
This operator ordering is obtained by placing the support of these operators on successive time slices.  
Since $U$ is topological, from a spacetime point of view, the same result obtains if instead we deform 
the surfaces $\Sigma$ and $-\Sigma$ to a single surface $S$ in spacetime that {\it surrounds} the locus $C$, as illustrated here in cross-section:
\begin{equation}
 \parbox{.19\textwidth}{
\begin{tikzpicture}
	\draw[ decoration={markings, mark=at position 1 with {\arrow{>}}},
        postaction={decorate}] (-1.2, -.4)--(-1.2, .4); 
        \node (A) at (-1, 0) {$t$};
        \node[] (A) at (.7, -.3) {$-\Sigma$};
        \node[darkblue] (A) at (-.25, 0) {$C$};
        \node (A) at (.8, .3) {$\Sigma$};
        
	\draw[thick,    decoration={markings, mark=at position 0.5 with {\arrow{<}}},
        postaction={decorate}] (-1,.5) -- (1,.5); 
	\draw[thick,   decoration={markings, mark=at position 0.5 with {\arrow{>}}},
        postaction={decorate}] (-1,-.5) -- (1,-.5); 
	\filldraw[darkblue, fill=darkblue] (0,0) circle (.07); 	
\end{tikzpicture}
}
= ~
\parbox{.2\textwidth}{
\begin{tikzpicture}
	\filldraw[darkblue, fill=darkblue] (0,0) circle (.07); 	
	\draw[thick,  
        decoration={markings, mark=at position 0.125 with {\arrow{>}}},
        postaction={decorate}] (0,0) circle (.5);
        \node[darkblue] (A) at (-.25, 0) {$C$};
        \node (A) at (.7, 0) {$S$};
       
\end{tikzpicture}
}
\end{equation}
  The variation of the operator 
then depends on the {\it linking number} of $S$ and $C$ in spacetime.}.

In the case of a discrete 1-form symmetry, there is no current,
but the symmetry operator $U_\alpha(\Sigma)$ is still topological.  
If the 1-form symmetry group is $\IZ_p$, strings can disappear or end only in groups of $p$.

For general integer $p \geq -1$, a $p$-form symmetry means the existence of topological operators $U_\alpha(\Sigma_{D-p-1})$ 
labelled by a group element $\alpha$ and a closed codimension-$(p+1)$ submanifold of spacetime\footnote{For discussion of $p=-1$, see 
\cite{Cordova:2019jnf}.}.  
For coincident submanifolds, these operators satisfy the ``fusion rule"
$ U_\alpha(\Sigma) U_\beta(\Sigma) = U_{\alpha+ \beta} (\Sigma)$.  
The operators charged under a $p$-form symmetry are supported on $p$-dimensional loci, and create $p$-brane excitations.  
The conservation law asserts that the $(p+1)$-dimensional worldvolume of these excitations will not have boundaries.

For $p\geq 1$, the symmetry operators commute with each other -- higher-form symmetries are abelian \cite{Gaiotto:2014kfa}.
To see this, consider a path integral representation of an expectation value with two symmetry operators 
$U(\Sigma_1) U(\Sigma_2)$ inserted on the same time slice $t$.   The ordering of the operators can be specified in the path integral by shifting the left one to a slightly later time $t+\epsilon$.  If $p\geq 1$, then $\Sigma_{1,2}$ have codimension larger than one, and their locations can be continuously deformed to reverse their order.  

\subsection{Physics examples of higher-form symmetries}

\begin{itemize}

\item Maxwell theory in $D=3+1$ with electric charges but no magnetic charges has a continuous 1-form symmetry with current
$J_{(m)}^{\mu\nu} = {1\over 4 \pi} \epsilon^{\mu\nu\rho\sigma} F_{\rho \sigma}\equiv {1\over 2 \pi}( d\tilde A)^{\mu\nu}$.
The statement that this current is conserved $\nabla_\mu J^{\mu\nu}_{(m)}=0$ is the Bianchi identity expressing the absence of magnetic charge.
The symmetry operator is 
$ U_\alpha^{(m)}(\Sigma) = e^{ { \ii \alpha \over 2\pi } \int_\Sigma F } $.  
The fact that the charge operator $ \int_\Sigma F$ depends only on the topological class of $\Sigma$ is the magnetic gauss law -- 
when $\Sigma$ is contractible, it counts the number of magnetic monopoles inside.  
This symmetry shifts the dual gauge field $\tilde A$ by a flat connection;
the charged line operator is the 
't Hooft line, $W^{(m)}[C] = e^{ \ii \oint_C \tilde A}$.

In free Maxwell theory without electric charges,
there is a second 1-form current,
$J_{(e)}= F$ 
whose charged operator is the Wegner-Wilson line $W^{(e)}[C] = e^{ \ii \oint_C  A}$.
The symmetry operator for this 
`electric' 1-form symmetry is $ U_{\alpha}^{(e)} (\Sigma_2) = e^{ \ii  { 2 \alpha \over g^2} \int_{\Sigma_2} \star F}$, which
(by canonical commutators) shifts the gauge field $A$ by a flat connection.

\item
Pure $\gSU(N)$ gauge theory or $\IZ_N$ gauge theory or $\gU(1)$ gauge theory with charge-$N$ matter 
has a $\IZ_N$ 1-form symmetry, called the `center symmetry'.
The charged line operator is the Wegner-Wilson line in the minimal irrep, $W[C] = \tr P e^{\ii \oint_C A } $.

\item 
When we spontaneously break a 0-form U(1) symmetry in $d=2$, 
there is an emergent 1-form U(1) symmetry
whose charge counts the winding number of the phase variable $\varphi$
around an arbitrary closed loop $C$,
$Q[C] = \oint_C d \varphi $.
In $d$ spatial dimensions, this is a $(d-1)$-form symmetry.
The charged operator creates a vortex (in $d=2$, or a vortex line or sheet in $d>2$).
Unlike the examples above, this symmetry is generally not an exact symmetry of a microscopic Hamiltonian for a superfluid;
it is explicitly broken by the presence of vortex configurations.
More on this example and its consequences for superfluid physics in \S\ref{sec:anomalies-of-higher-form}.

\item There is a sense in which the 3d Ising model has a $\IZ_2$ 1-form symmetry reflecting the integrity of domain walls between regions of up spins and regions of down spins.
The charged line operator is the Kadanoff-Ceva disorder line \cite{kadanoff1971determination} -- the boundary of a locus along which the sign of the Ising interaction is reversed (for a review, see 
\cite{Fradkin:2016ksx}).  But because a domain wall is always the boundary of some region, no states are charged;
relatedly, the disorder line is not a local string operator.  
If we gauge the $\IZ_2$ symmetry of the Ising model, the disorder line becomes the Wegner-Wilson line of the resulting $\IZ_2$ gauge theory, and this theory has a genuine 1-form symmetry.


\end{itemize}

\label{subsec:SF}

\subsection{Spontaneous symmetry breaking}

Anything we can do with ordinary (0-form) symmetries, we can do with higher-form symmetries.
In particular, they can be spontaneously broken.

One way to characterize the unbroken phase of a 0-form symmetry is 
that correlations of charged operators are short-ranged, meaning that 
they decay exponentially with the separation between the operators
\be  \vev{ \CO(x)^\dagger \CO(0)}  \sim e^{ - m |x| } .\ee
A language that will generalize is to regard the two points at which we insert 
a charged operator and its conjugate as an $S^0$, a zero-dimensional sphere, 
and the separation between the points as the size of the sphere.
The broken phase for 0-form symmetry can be diagnosed by long-range correlations:
\be \vev{ \CO(x)^\dagger \CO(0)}  = \vev{\CO^\dagger(x)} \vev{\CO^{\nd}(0)}+ \cdots 
~,\ee
independent of the size of the $S^0$.

For a $p$-form symmetry, the unbroken phase is also when correlations of charged operators are short-ranged, 
and decay when the charged object grows.  
For a 1-form symmetry, this is when the charged loop operator exhibits an area law:
\be \vev{ W(C) }  \sim e^{ - T_{p+1} \text{Area}(C) } ,\ee
where $\text{Area}(C)$ is the area of the minimal surface bounded by the curve $C$.
In the case of electricity and magnetism, an area law for $\vev{ W^E(C) }$ is the superconducting phase.

The broken phase for a $p$-form symmetry is signalled by a failure 
of the expectation value of the charged operator to decay with size.  
For a 1-form symmetry, this is when the charged loop operator exhibits a perimeter law:
\be \label{eq:perimeter-law} \vev{ W(C) }  = e^{ - T_{p} \text{Perimeter}(C) }  + \cdots .\ee
The coefficient $T_p$ can be set to $0$ by modifying the definition of $W(C)$ by 
counterterms local to $C$, 
so \eqref{eq:perimeter-law} says that a large loop has an expectation value.

The fact that charged operators have long-range correlations means that the generators of the symmetry 
act nontrivially on the groundstate.  
More precisely, $\ket{\psi}$ is not stationary under the symmetry (SSB)
 if and only if 
there exists   
a charged operator $O$
with $\bra{\psi} O \ket{\psi} \neq 0 $ (long-range order).
So, LRO $\Leftrightarrow$ SSB.
I put a proof of this statement in Appendix \ref{App:SSB-LRO}.

SSB of higher-form symmetry has been a fruitful idea. 
 In the next two subsections, I'll illustrate the consequences in the case of discrete and continuous symmetries, respectively.

\subsection{Topological order as SSB}

One definition of topological order is the presence of a groundstate subspace of {\it locally indistinguishable} states, as in \eqref{eq:indistinguishable}.
This means that no local operator takes one groundstate to another; instead the operator that takes one groundstate to another is necessarily an extended operator.  
But this is a description of 
the spontaneous breaking of a higher-form symmetry \cite{Gaiotto:2014kfa, Wen:2018zux}: 
the generators of the broken part of the higher-form symmetry commute with the Hamiltonian (at least at low energies) and act nontrivially on the groundstates\footnote
{Addendum in v3: However, SSB of higher-form symmetry by itself is not sufficient to imply topological order \cite{Huxford:2023bhb}.
The loophole is the following.  Just because two groundstates are related by the action of a nonlocal operator does not mean that there isn't some linear combination of them that are related by a local operator.  
And in fact there is a counterexample, a state with SSB of a 1-form symmetry which does not have topological order \cite{castelnovo2007quantum,Huxford:2023bhb}.  I have added an appendix \ref{App:chamon-state} explaining some details of this counterexample.
In the discussion of the toric code below, the extra assumption is that the charged operators $V(C)$ are also topological.  Alternatively, we could describe the extra assumption as requiring SSB of an {\it anomalous} higher-form symmetry.  
}.

Let's think about the example of $\IZ_p$ gauge theory (whose solvable limit is the toric code \cite{kitaev1997}) in $D$ spacetime dimensions.  
This is a system with $\IZ_p$ 1-form symmetry
with symmetry operators $U(M_{D-2})$, supported on a $(D-2)$-dimensional manifold,
and charged operators $V(C_1)$, supported on a curve.  
In terms of the toric code variables, we can be completely explicit.  
On each link we have a $p$-state system on which act the Pauli operators $Z = \sum_{k=0}^p \omega^k \ketbra{k}{k}$ and $X = \sum_{k=0}^p \ketbra{k+1}{k}$ 
(where $\omega \equiv e^{ 2\pi \ii/p}$ and  the arguments of the kets are understood mod $p$).
Then
 $V(C) = \prod_{\ell \in C} X_\ell$, $U(M) = \prod_{\ell \perp M} Z_\ell $,
 where we regard $M$ as a surface in the dual lattice, and $ \ell \perp M$ indicates all links crossed by the surface $M$.
Their algebra is
\be \label{eq:Z_p-gauge-theory-algebra} U^m(M) V^n(C) = e^{ 2 \pi  \ii {m n \over p} \#(C, M) } V^n(C) U^m(M) \ee
where 
$\#(C, M)$ is the intersection number of the curve $C$ with the surface $M$.
This is the algebra of electric strings and magnetic flux surfaces in $\IZ_p$ gauge theory.
Deep in this gapped phase, $H=0$, and there is a description in terms of topological field theory.
A simple realization is $BF$ theory of a 1-form potential $a$ and $(D-2)$-form potential $b$, with action 
\be S[b,a] = { p \over 2 \pi} \int_{D} b \wedge da  
= { p \over 2 \pi } \int d^D x \epsilon^{\mu_1 \cdots \mu_D} b_{\mu_1 \cdots \mu_{D-2} } \partial_{\mu_{D-1} } a_{\mu_D}
\ee
in terms of which 
\be U^n(M ) = e^{ \ii n \oint_M b_{D-2} },~~~V^m(C) = e^{ \ii m \oint_C a }  .\ee
The algebra \eqref{eq:Z_p-gauge-theory-algebra} follows from canonical commutation relations in this gaussian theory.
Since $V(C)$ has a perimeter law in the deconfined phase, the charged objects whose condensation breaks the 1-form symmetry are the lines of electric flux.

Another example is the Laughlin fractional quantum Hall states.  
So far in our discussion the symmetry operators for a 1-form symmetry with group $A$ form a representation of $A$
on the 1-cycles of space, $Z$, {\it i.e.}~a linear map 
$U: Z \to U(1)$,
where the representation operators commute
$U(M)U(M') = U(M')U(M)$.
This relation can be generalized to allow for phases -- i.e. a projective representation.  
Consider a system in $D=2+1$ with a $\IZ_k$ 1-form symmetry
that is realized projectively in the following sense:  
\be \label{eq:projective-laughlin} U^m(C) U^n(C') = e^{ { 2 \pi  \ii m n \#(C, C') \over k} } U^n(C') U^m(C)~~~\ee
where $\#(C,C')$ is the intersection number of the two curves $C, C'$ in space.
Regarding $U(C)$ as the holonomy of a charged particle along the loop $C$,
this is the statement that flux carries charge.
Representing this algebra nontrivially gives $k$ groundstates on $T^2$.
This algebra, too, has a simple realization via abelian Chern-Simons theory, $S[a] = { k \over 4 \pi} \int a \wedge da $,
with $U^m(C) = e^{ \ii m \oint_Ca }$.  

The algebra in \eqref{eq:projective-laughlin} is a further generalization of 1-form symmetry, in that the group law is only satisfied up to a phase.  
As we will discuss in \S\ref{sec:anomalies}, it is an example of a 1-form symmetry  anomaly.

The preceding discussion applies to {\it abelian} topological orders.
In this context, abelian means that 
the algebra of the line operators transporting the anyons forms a group, which must be abelian by the argument above.
In \S\ref{sec:cat} we discuss the further generalization that incorporates non-abelian topological orders.

\subsection{Photon as Goldstone boson}

What protects the masslessness of the photon?  The case of quantum electrodynamics (QED) is the most visible version of this question; the same question arises in condensed matter 
as: why are there $\gU(1)$ spin liquid phases, with an emergent photon mode?

Higher-form symmetries provide a satisfying answer to this question (unlike appeals to gauge invariance, which is an artifact of a particular description): the gaplessness of the photon can be understood as required by spontaneously-broken $\gU(1)$ 1-form symmetry
\cite{Kovner:1992pu, Gaiotto:2014kfa,Hofman:2018lfz, Lake:2018dqm}, as a generalization of the Goldstone phenomenon.

Here is a perspective on the zero-form version of the Goldstone theorem.  
Given a continuous zero-form symmetry with current $j_\mu$, we can couple 
to a background gauge field ${\cal A}$  by adding to the Lagrangian $ \Delta L \ni j_\mu {\cal A}^\mu$.
If the symmetry is spontaneously broken, the effective Lagrangian will contain a Meissner term proportional to $\CA^2$.
But the effective action must be gauge invariant, and this requires the presence of a field that transforms 
nonlinearly under the $\gU(1)$ symmetry: $\varphi \to \varphi + \lambda, \CA \to \CA - d \lambda$;
this is a global symmetry if $d\lambda=0$.
Altogether, the effective Lagrangian must contain a term of the form
\be\CL_\text{eff} = -{1\over 4\pi g} \(d \varphi + \CA \)^2 ~\label{eq:goldstone-action}\ee
(where by $(\omega)^2$ I mean 
$\omega_p \wedge \star \omega_p  
= {1\over p!}  \omega_{\mu_1 \cdots \mu_p} \omega^{\mu_1 \cdots \mu_p} $).
The coefficient $ {1\over 4 \pi g}$ is the superfluid stiffness.  

The analog for a continuous 1-form symmetry works as follows.  
The current is now a two-form, so the background field must be a two-form gauge field $\CB_{\mu\nu}$ 
and the coupling is 
$ \Delta L \ni J_{\mu\nu} \CB^{\mu\nu}$.
The same logic implies that the effective action for the broken phase must contain a term 
\be\CL_\text{eff} = -{1\over 2 g^2 } \(d a + \CB \)^2 \label{eq:1-form-goldstone-action} \ee
where the Goldstone mode $a$ is a 1-form
that transforms nonlinearly $ a \to a + \lambda, \CB \to \CB - d \lambda$;
this is a global symmetry if $d\lambda =0$.
Setting the background field $\CB=0$, we recognize this as
a Maxwell term for $a$. 
The coupling strength $g$ is determined by the analog of the superfluid stiffness.

For $p$-form $\gU(1)$ symmetry, we conclude by the same logic that there is a massless $p$-form field $a$ 
with canonical kinetic term 
\be S_\text{Max}[a] =  - {1 \over 2g^2}  \int da \wedge \star da .\ee

Returning to QED, we see that the familiar Coulomb phase is the SSB phase for the $\gU(1)$ 1-form symmetry.  
The unbroken phase is the superconducting phase, where the photon has short-ranged correlations.
(In an ordinary superconductor, where the Cooper pair has charge two, a $\IZ_2$ subgroup of the 1-form symmetry remains broken.)

As in the case of 0-form SSB, the broken phase can be understood via the condensation of charged objects;
in this case the charged objects are the strings of electric flux \cite{Levin:2004mi, Levin:2004js}.
Notice that the presence of charged matter, on which these strings can end, and which therefore explicitly breaks this symmetry, does not necessarily destroy the phase.  We'll comment on this robustness more in \S\ref{subsec:robustness}.
In fact, because of electromagnetic duality, the Coulomb phase is the broken phase for {\it either} the electric 1-form symmetry or the magnetic 1-form symmetry \cite{Gaiotto:2014kfa}.

\subsection{Effects of IR fluctuations}
The analog of the Hohenberg-Coleman-Mermin-Wagner (HCMW) theorem
for higher-form symmetries 
\cite{Nussinov:2006iva, Nussinov:2009zz, Gaiotto:2014kfa, Lake:2018dqm}
is interesting.  
As in the proof of the HCMW theorem, we suppose that a $p$-form $\gU(1)$ symmetry in $D$ spacetime dimensions is spontaneously broken
and that there is therefore a Goldstone mode, a massless $p$-form field $a$.
Then we ask if indeed the symmetry is broken by evaluating the 
expectation value of a charged operator $W_C$, including the fluctuations of the would-be-Goldstone mode $a$.
We can choose $C$ to be a copy of $\IR^p \subset \IR^D$ so that we can do the integrals, and the result is 
(see \cite{Lake:2018dqm} for a discussion of a convenient gauge choice)
\be 
\vev{W_C} = 
Z^{-1}\int [Da] e^{ - S_\text{Max}[a] + \ii \int_C a } \simeq 
\exp{ \(  - \half g^2 L^p \int {\dbar^{D-p} k \over k_\perp^2 }  \) } 
\label{eq:wilson-vev}\ee
where $\dbar k \equiv { dk \over 2\pi}$ and $k_\perp$ is the momentum transverse to $C$.
The integral in the exponent of \eqref{eq:wilson-vev} is IR divergent when $D-p \leq 2$.
As in the $p=0$ case, we interpret this as the statement that the long-wavelength fluctuations 
of the would-be-Goldstone mode necessarily destroy the order.
(For $D-p \geq 2$, the integral is UV divergent.  This divergence can be absorbed in a counterterm locally redefining the operator 
$W_C \to W_C e^{ - \delta T \int_C d^p x} $,
which can be interpreted as a renormalization of the tension $T$ of the charged brane.)
In the marginal case of $p = D-2$, the long-range order is destroyed, but $\vev{W_C}$ decays as a power-law in the loop size, rather than an exponential; 
this is a higher-form analog of algebraic long-range order in $D=2$.  

The calculation above is independent of compactness properties of the Goldstone form field,
in the sense that in \eqref{eq:wilson-vev} we just did a Gaussian integral over the topologically-trivial fluctuations of $a$.
In the marginal case $D=2+1, p=1$, 
if we treat $a$ as a compact $\gU(1)$ gauge field, SSB of the 1-form symmetry is avoided instead because
monopole instantons generate a potential for the dual photon $d \sigma = \star d a/2\pi$ \cite{P7729}.
This mechanism generalizes to any case with $D-p=2$ \cite{Lake:2018dqm}.

\cite{Nussinov:2006iva, Nussinov:2009zz} interpret such results as a generalization of Elitzur's theorem on the unbreakability of local gauge invariance \cite{PhysRevD.12.3978}.

\subsection{Robustness of higher-form symmetries}
\label{subsec:robustness}

We are used to the idea that consequences of emergent (aka accidental) symmetries are only approximate:
explicitly breaking a spontaneously-broken continuous 0-form symmetry gives a mass to the Goldstone boson.

This raises a natural question.
The existence of magnetic monopoles with $m = M_\text{monopole}$ explicitly breaks the 1-form symmetry of electrodynamics:
$$ \partial^\mu J_{\mu\nu}^{E} = j^\text{monopole}_\nu~.$$
If the photon is a Goldstone for this symmetry, does this mean the photon gets a mass?
Perhaps surprisingly, the answer is `no' (early discussions of the robustness of broken higher-form symmetries using different words
include \cite{Forster:1980dg, Wegner1971, hastings2005quasiadiabatic}).  This is a way in which zero-form and higher-form symmetries are quite distinct.

A cheap way to see that `no' is the right answer is by dimensional analysis.
How does the mass of the photon $m_\gamma$ depend on the mass of the magnetic monopole, $M_\text{monopole}$?  
Suppose all the electrically charged matter (such as the electron) is very heavy or massless.
We must have $m_\gamma \to 0$ when $M_\text{monopole} \to \infty$.  But there is no other mass in the problem to make up the dimensions.

A slightly less cheap way to arrive at this answer is by dimensional reduction.   
If we put quantum electrodynamics (QED) on a circle of radius $R$, we arrive at low energies at abelian gauge theory in $D=2+1$,
which is confined by monopole instantons \cite{P7729}.   
The monopole instantons arise from euclidean worldlines of magnetic monopoles wrapping the circle, 
and so their contribution to the mass of the (2+1)d photon is
\be m_\gamma(R) \sim e^{ - R M_\text{monopole} }. \ee
The polarization of the photon along the circle gets a mass from euclidean worldlines of charged matter (like the electron) wrapping the circle, so
its mass is
\be  m_4(R) \sim e^{ - R m_e} . \ee
But now the point is simply that when $R\to \infty$, both of these effects go away and the (3+1)d photon is massless. 

A third argument is that operators 
charged under a 1-form symmetry are loop operators -- they are not local.
We can't add non-local operators to the action at all.
This argument is not entirely satisfying, since on the lattice even the action for pure gauge theory is a sum over (small) loop operators.  
The question is whether the dominant contributors in this ensemble of charged loop operators grow under the RG.
\cite{Iqbal:2021rkn} describes a toy calculation to address this question: 
begin in a phase with a perimeter law $\vev{W[C]} \sim t^{\text{length}[C]}$ and 
consider adding to the action $ g \int [dC] W[C] + h.c. $ in perturbation theory in $g$.  
Regularizing on the lattice and neglecting collisions of loops, the result is the same as integrating out a charged particle 
whose mass is determined by the parameter $t$.  For small enough $t$ there is an IR divergence indicating a transition to a phase where the charged particle is condensed. 
Until that happens, the SSB phase survives.
A useful slogan extracted from this calculation is that a loop operator becoming relevant (changing the IR physics) indicates the onset of a Higgs transition.

The discrete analog of this phenomenon is instructive.  
In the solvable toric code model, the discrete 1-form symmetries are exact. 
But in the rest of the deconfined (spontaneously broken) phase, they are {\it emergent}, but still spontaneously broken,
and still imply a topology-dependent groundstate degeneracy that becomes exact in the thermodynamic limit.
A rigorous proof of this \cite{hastings2005quasiadiabatic} constructs (slightly thickened) string operators by 
quasi-adiabatic continuation.

Known forms of topological order in $d\leq 3+1$ have the property that at any $T>0$
they are smoothly connected to $T=\infty$ (a trivial product state).
If the 1-form symmetry is emergent, then as soon as $T>0$, a mass {\it is} generated for 
the photon (by the argument above, with the circle regarded as the thermal circle, so that $R=1/T$), and the state is smoothly connected to $T=\infty$.  

We do know an example of a topologically ordered phase that is stable at $T>0$, 
namely the two-form toric code in $D=4+1$ \cite{Dennis:2001nw}.  
In the $\gU(1)$ version of this theory, 
the masslessness of the two-form gauge field should survive explicit short-distance breaking of the $\gU(1)$ two-form symmetry, even at finite temperature.
The reason is that a theory with a two-form symmetry on a circle still has a 1-form symmetry.  

We conclude that the consequences of higher-form symmetries are more robust to explicit breaking than zero-form symmetries.

\subsection{Mean field theory}

Landau-Ginzburg mean field theory is our zeroth-order tool for understanding symmetry-breaking phases and their neighbors.
It is therefore natural to ask whether it has an analog for higher-form symmetries \cite{Iqbal:2021rkn}.
We focus on the simplest nontrivial case of a $\gU(1)$ 1-form symmetry.

It is worthwhile to review the logic
that produces this weapon.  
If we take the Landau paradigm seriously, then the only low-energy modes we require are those swept out by the symmetry.  
The key idea is to introduce a degree of freedom $\phi(x)$ at each point in space that transforms {\it linearly} under the symmetry.
$\phi$ should be regarded as a coarse-grained object, and this is an effective long-wavelength description.  
In the example of a magnet, $\phi(x)$ can be the magnetization averaged over a small cell at $x$.
Now, because there are no other light degrees of freedom (by assertion \ref{item1}), the effective action for $\phi$ should be given by an analytic functional of $\phi$ which is local in spacetime.  This functional can therefore be expanded in a series consisting of all symmetric local functionals of $\phi$, organized in a derivative expansion of terms of decreasing relevance.  The length scale suppressing higher derivates is the short distance over which we averaged in constructing $\phi(x)$.

The 1-form analog of the order parameter field $\phi(x)$ (which is a function from the space of points into a linear representation of $G$), 
is a functional $\psi[C]$ from the space of loops into a linear representation of $G$, a `string field'.  
While $\phi(x)$ transforms under the zero-form symmetry as $\phi(x) \to \phi(x) e^{ \ii \alpha }$, with $d\alpha =0$,
the 1-form analog transforms like $ \psi[C] \to \psi[C] e^{ \ii \oint_C \Gamma}$, with $d \Gamma =0$.

\begin{figure}[h!]
$$\parfig{.4}{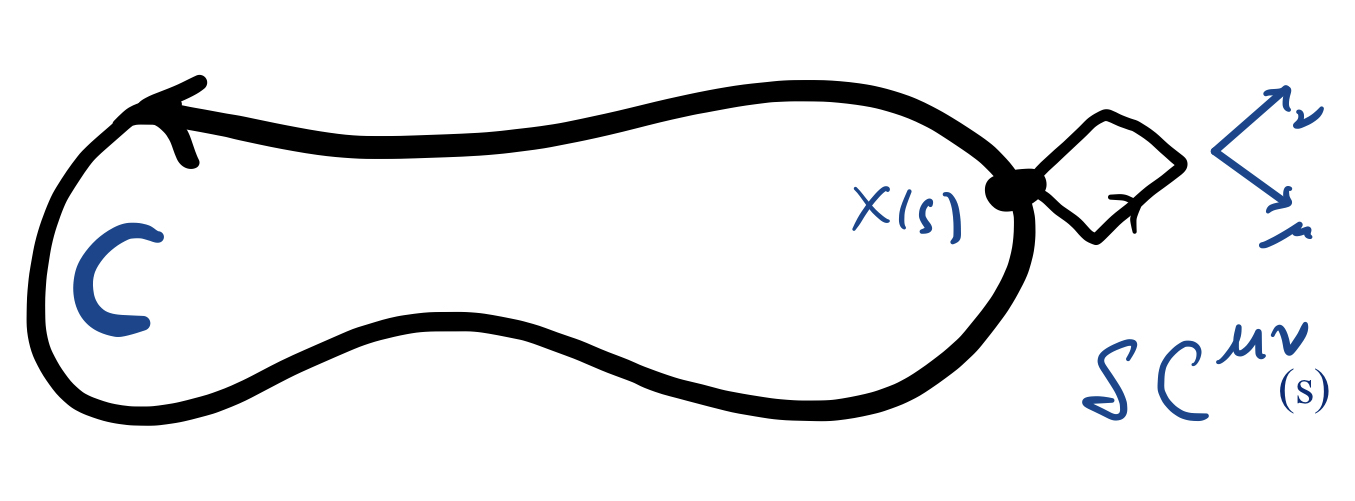}
~~\parfig{.4}{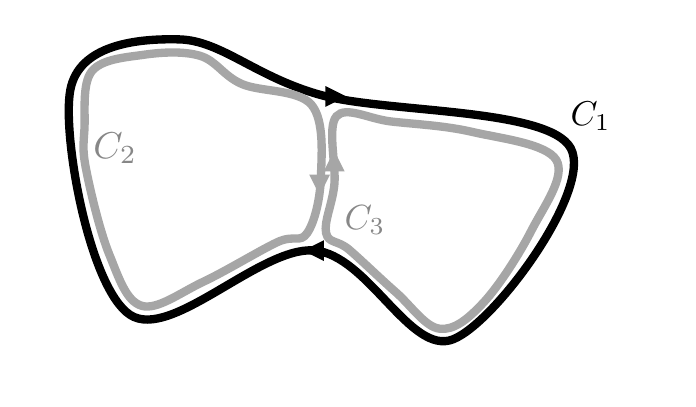}$$
\caption{Left: a sketch of the definition of the area derivative $ { \delta \over \delta C_{\mu\nu}(s)}$. 
Right: The arrangement of loops involved in the topology-changing term  $S_r$ in the MSFT action.}
\label{fig:area-derivative}
\end{figure}

To write an action for such a field requires the analog of a derivative, which compares its values on nearby loops.
Such an `area derivative' was discovered in the study of loop-space formulations of gauge theory \cite{migdal1983loop} (see Fig.~\ref{fig:area-derivative}, left).
The analog of integrating the action over spacetime $\int d^Dx $ is integrating over the space of loops $\int [dC]$.  
The most general action consistent with the symmetries then takes the form
\be S[\psi] = \int [dC] \( V(|\psi[C]|^2 ) + { 1\over 2 L[C] } \oint ds 
{ \delta \psi^\star[C] \over \delta C_{\mu\nu}(s) } 
{ \delta \psi[C] \over \delta C^{\mu\nu}(s) }  + \cdots  \) + S_r[\psi] ~.
\label{eq:MSFT}\ee
The last `recombination term' 
\be S_r[\psi] =  \int [dC_{1,2,3}] \delta[C_1 - (C_2 + C_3) ] \( \lambda \psi[C_1] \psi^\star [C_2]\psi^\star[C_3] + h.c. \) + \cdots \ee
is not local in loop space, but is local in real space since it involves only a single integral over the center-of-mass of the loops.
Here the delta function imposes the equality of loops regarded as integration domains  (see Fig.~\ref{fig:area-derivative}, right).
The $\cdots$ denote terms with more derivatives or more powers of $\psi$.
Models similar to this Mean String Field Theory (MSFT) have been considered before in various specific contexts \cite{Banks:1980sq, Yoneya:1980bw, Rey:1989ti, Franz:2006gb, Beekman_2011}.

The potential term $V(|\psi[C]|^2 ) = r |\psi[C]|^2 + u |\psi[C]|^4 + \cdots $ controls the low-energy behavior. 
If $r>0$, we find an unbroken phase where $\psi[C] \simeq e^{ - \sqrt{r} A[C]} $.  
When $r<0$, the strings want to condense.  The fluctuations around nonzero $|\psi|$ are all massive, except for the 
geometric mode $ \psi[C] = v e^{ \ii \oint_C ds a_\mu(x(s)) \dot x^\mu (s) } $, 
which describes a slowly-varying 1-form symmetry transformation,
and in terms of which the action \eqref{eq:MSFT} reduces to the Maxwell action for $a$, with coupling $g^2 = {1\over 2v^2}$.

As in the zero-form case, another application of this mean field theory is to classify topological defects of the resulting ordered media.  
The conclusion for $G = \gU(1)$ is that the only topological defect is the codimension-three magnetic monopole.

So far, we have discussed the case of a $\gU(1)$ 1-form symmetry.  
The case of discrete symmetries can be approached by 
 explicitly breaking the $\gU(1)$ 1-form symmetry to a discrete subgroup. 
 A term of the form 
 \be\label{eq:Z-k-breaking} h \int [dC] \psi^k[C] + {\rm h.c.} \ee 
 breaks it down to $\IZ_k$.
 In the broken phase, the effective action reduces to a continuum (BF) description of $\IZ_k$ gauge theory.

The action \eqref{eq:MSFT} can be given a lattice definition and contact can be made with microscopic Hamiltonians in the following way.  
Zero-form mean field theory arises from a variational 
using a product state ansatz $\ket{\Psi_\phi} = \otimes_x \ket{\phi(x)}$; 
given a microscopic Hamiltonian, 
the variational energy $\bra{\Psi_\phi } \HH  \ket{\Psi_\phi} = H[\phi]$ takes the form of the Landau-Ginzburg Hamiltonian.  

Consider for definiteness a $\IZ_2$ lattice gauge theory Hamiltonian, in the form 
\be H_\text{TC} = - \infty \sum_{\text{sites}, s} A_s - \Gamma \sum_{\text{plaquettes}, p} B_p - g \sum_{ \text{links}, \ell} Z_\ell .\ee
This acts on a Hilbert space that is a tensor product of qubits on the links of a cell complex; 
$A_s = \prod_{\ell \in s} Z_\ell$ and $B_p = \prod_{\ell \in p} X_\ell$.  $X$ and $Z$ denote the Pauli operators.
In the $Z$ eigenbasis, we regard a link as covered by a segment of string if $Z=-1$.
We take the coefficient of the star term $A_s$ to infinity so that the loops are closed and there is an exact (electric) 1-form symmetry generated by 
$U(C) =  \prod_{\ell \in C} X_\ell$.
When $g=0$, the groundstate is the uniform superposition over all collections of closed loops.  
$g$ represents a tension for the electric strings; for large enough $g/\Gamma$, there is a transition to a confined phase.   

The analog of a product state for the 1-form case is a many-body wavefunction on collections of loops determined by a function $\psi[C]$ on a single loop:
\be \ket{\Psi_\psi} = : e^{ \sum_{c, \text{connected}} \psi[c] U[c] }: \ket{0} \ee
where $U[c] \ket{0} = \ket{c}$ creates the loop $c$, and the normal-ordering symbol $: \cdots :$ is a prescription for dealing with overlapping loops.  
The variational energy for this state is a lattice Hamiltonian for the action \eqref{eq:MSFT} plus \eqref{eq:Z-k-breaking}.

{\bf Brief comments on phase transitions.}  As in the 0-form case, we expect the mean-field description to break down near critical points, below the upper critical dimension.  (The extension of the renormalization group to MSFT has not yet been attempted.)  Dimensional analysis says that the string field $\psi$ has dimension $(D-4)/2$ and hence $u$ has mass dimension $8-D$, and $\lambda$ has dimension $ 6-D/2$, which puts the upper critical dimension at 8 or 12, depending on which coupling matters.  More significantly, the recombination term is a symmetric term {\it cubic} in the order parameter field, and we expect that it generically renders the transition first order. This is consistent with numerical work on deconfinement transitions in gauge theory in $D>3$ (see \eg~\cite{Florio:2021uoz} and references therein).   

Notice that the string field has engineering dimension zero in $D=4$.  There are two possible notions of lower critical dimension, which coincide at $D=2$ in the case of 0-form symmetries.  One is the largest dimension where the HCMW theorem forbids symmetry breaking, which is $D=3$ for 1-form symmetry.  
The other is the dimension in which the linearly-transforming field is classically dimensionless, which is $D=4$ for 1-form symmetries.  
In the case of 0-form symmetry, this allows for the rich physics of the Berezinsky-Kosterlitz-Thouless transition, where 
there is a line of (free) fixed points (parameterized by $g$ in \eqref{eq:goldstone-action}) that terminates when a symmetry-allowed operator becomes relevant.  
A universal prediction is the value of the stiffness at the transition, since as in \eqref{eq:goldstone-action}, the stiffness determines the coupling.

For the special case of 1-form symmetry-breaking in $D=4$, there is again a line of (free) fixed points, parameterized by the Maxwell coupling,
as in \eqref{eq:1-form-goldstone-action}.  Consider the application of this picture to $\IZ_k$ gauge theory, described by perturbing the MSFT action by 
\eqref{eq:Z-k-breaking}.   In the free theory, this operator can be argued \cite{Kapustin:2005py} 
to have an anomalous dimension $ \Delta_k(g) = { g^2 k^2 \over 32\pi^2} $;
for large enough $k$, $\Delta_p(g)$ passes through $4$ at some $g_c < \sqrt{4\pi}$, and we can interpret this as the location of a transition 
in the low-energy physics from a Coulomb phase to a phase with $\IZ_k$ topological order.  
The prediction is again a universal jump in the `superfluid stiffness', namely the value of the gauge coupling at the transition.  
Many of these ideas were anticipated by Cardy \cite{Cardy:1980jg} without the benefit of the language higher-form symmetry.

There is a catch: this transition is observed in Monte Carlo simulations 
to actually be weakly first order (see \cite{Vettorazzo:2003fg} and references therein).
Does that mean there is nothing universal to say?  There is a reason the transition is {\it weakly} first order.  The magnetic charge whose condensation drives the transition has a good dual description via the Abelian Higgs model.  In this model, fluctuations drive the transition first order \cite{Coleman:1973jx}.  If the coupling is weak at the transition, this description is good and the transition is weakly first order.  
But, using the slogan of \S\ref{subsec:robustness} (a loop operator becoming relevant means a Higgs transition),
the mechanism of the previous paragraph determines the critical coupling and shows that it should be small at large $k$.

\section{Anomalies}
\label{sec:anomalies}

My motivation for including a discussion of anomalies here is twofold.  One is that anomalies are a necessary ingredient in a suitably-generalized Landau Paradigm that incorporates all phases, in particular topological insulators and SPT phases.  A second motivation is that, as I will review, the existence of anomalies makes symmetries much more useful for constraining the dynamics of a physical system, and their generalization to higher-form symmetries is therefore an essential step.

The historical, high-energy-physics perspective on anomalies starts from specifying a quantum field theory by a path integral 
\be Z = \int [D(\text{fields})] e^{ \ii S[\text{fields}]} .\ee
An anomaly is a symmetry of the action $S$ that is not a symmetry of the path-integral measure.
The first example found was the chiral anomaly, the violation of the axial current of a charged Dirac field (the symmetry that rotates 
left-handed and right-handed fermions with opposite phases)
\be\label{eq:chiral-anomaly} \partial_\mu j_A^\mu = N { e^2 \over 16 \pi^2 } \epsilon_{\mu\nu\rho\sigma}F^{\mu\nu} F^{\rho \sigma},\ee
which controls the decay of the neutral pion into two photons.

\begin{figure}[h!]
$$    \parfig{.45}{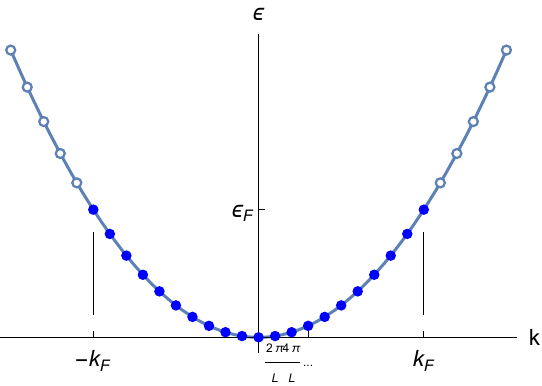}
    \parfig{.45}{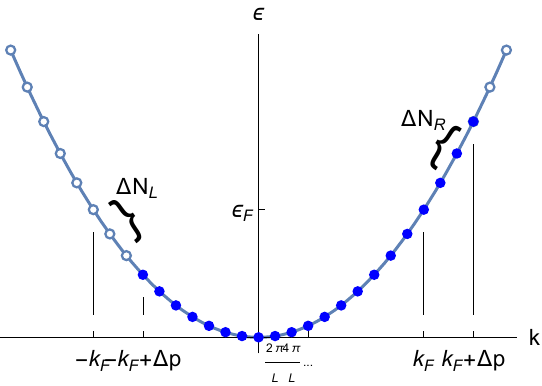}
$$
\caption{Left: Spectrum of a free-fermion tight-binding model in one dimension, near the bottom of the band at some small filling.  Green circles indicate filled states.  Right: The result of adiabatically applying an electric field.  $N_{L/R}$ indicate the number of left-moving and right-moving excitations.}
\label{fig:anomaly}
\end{figure}    

A more concrete perspective arises if we consider the same kind of system on the lattice, in one dimension for simplicity: consider a tight-binding model 
of fermions hopping on a chain, at some small filling as in Fig.~\ref{fig:anomaly}.
In this case, there is no chiral symmetry at all at the lattice scale.  It is an emergent symmetry, violated by the UV physics in a definite way.
At low energies, the system is approximately described by the neighborhood of the two boundaries of the Fermi sea, giving a 1d massless Dirac fermion, with a chiral symmetry.  
But if we adiabatically apply an electric field $E_x$, every fermion increases its momentum and the chiral charge changes by 
\be\label{eq:2danomaly} \Delta Q_A = \Delta (N_R - N_L) = 2 { \Delta p \over 2\pi/L}
={L\over \pi} e \int d t E_x (t) 
={ e\over 2\pi } \int \epsilon_{\mu\nu} F^{\mu\nu} .\ee
The left hand side is  $ \Delta Q_A = \int \partial^\mu j_\mu^A$,
and so \eqref{eq:2danomaly} is the 2d version of the chiral anomaly:
\be \partial_\mu j^\mu_A = { e\over 2 \pi} \epsilon_{\mu\nu}F^{\mu\nu}  .\ee

A reason for excitement about this phenomenon is that the coefficient $N$ in \eqref{eq:chiral-anomaly} is an integer.  
This is the first hint that an anomaly is a topological phenomenon, a quantity that is RG invariant \cite{'tHooft:1979bh}.
The idea is that the existence of the anomaly means that the partition function varies by some particular phase under the anomalous symmetry,
but an RG transformation must preserve the partition function.
Much of physics is about trying to match microscopic (UV) and long-wavelength (IR) descriptions.
That is, we are often faced with questions of the form
``what could be a microscopic Hamiltonian that produces these phenomena?"
and ``what does this microscopic Hamiltonian do at long wavelengths?". 
Anomalies are precious to us, because they are RG-invariant information:
any anomaly in the UV description must be realized somehow in the IR description.

Another useful perspective on anomaly is as an obstruction to gauging the symmetry.  
Gauging a symmetry means creating a new system where the symmetry is a redundancy of the description, by coupling to gauge fields.   
If the symmetry is not conserved in the presence of background gauge fields, the resulting theory would be inconsistent.

Above I've described an example of an anomaly of a continuous symmetry.  Discrete symmetries can also be anomalous.

Anomaly is actually a more basic notion than phase of matter:
The anomaly is a property of the degrees of freedom (of the Hilbert space) and how the symmetry acts on them, independent of a choice of Hamiltonian.  
Multiple phases of matter can carry the same anomaly.  

\subsection{SPT phases and anomalies}

The definition of gapped phases can be refined by studying only the space of Hamiltonians preserving some particular symmetry group $G$.  
Two phases that are distinct in this smaller space may nevertheless be connected by a gapped path in the larger space of non-symmetric Hamiltonians.

One way to define \cite{chen2013symmetry} a Symmetry-Protected Topological (SPT) phase is as a nontrivial phase of matter 
(with some symmetry $G$) without topological order  (for a review, see \cite{Senthil:2014ooa}).  
SPT phases can be characterized by their edge states.  
The idea is that the edge theory has to represent an anomaly of the symmetry $G$.  
It is really this anomaly that labels the bulk phase.  This phenomenon is called {\it anomaly inflow}.

As a simple example, consider an effective field theory for the integer quantum Hall effect, regarded as an SPT for charge conservation symmetry\footnote{Actually, the integer quantum Hall phase is more robust, and survives explicit breaking of the charge conservation symmetry.  It is protected by the gravitational anomaly manifested in the nonzero chiral central charge.}.
The charge conservation symmetry is associated, by Noether's theorem, with a conserved current $j^\mu$, with 
$ \partial_\mu j^\mu = 0$.
In $D=2+1$, this equation can be solved by writing 
$j^\mu = \epsilon^{\mu\nu\rho} \partial_\nu a_\rho/(2\pi)$,
in terms of a 1-form gauge field $a_\mu$, with redundancy $a \to a + d \alpha$.
The leading effective action for such a field, in the absence of parity symmetry, is 
a Chern-Simons term \cite{WZ9290,Zee:1996fe}:
\be S_\text{IQH}[a,\CA] = { 1\over 4 \pi } \int_{M} \epsilon^{\mu\nu\rho} \( a_\mu \partial_\nu a_\rho 
+ 2 \CA_\mu \partial_\nu a_\rho \) ~\ee
where $\CA$ is a background field for the charge conservation symmetry.
Under $\CA \to \CA + d \lambda$, $\delta S_\text{IQH} = \int_{\partial M} {\epsilon^{ij}  \over 4\pi} f_{ij} \lambda$.
This is the contribution to the chiral anomaly from a single right-moving edge mode.

In terms of the definition of the anomaly as a variation of the partition function of the edge theory in the presence of background fields,
the variation of the bulk action cancels the anomaly of the edge theory, so that the whole system is $G$ symmetric.
The edge theory cannot be trivial, since it has to cancel the variation of the bulk under the symmetry transformation: it has to be either  \cite{PhysRevX.3.011016}
\begin{itemize}
\item gapless 
\item symmetry-broken
\item or topologically ordered.
\end{itemize}
In particular, there cannot be a trivial gapped groundstate.
These are the same conditions arising from the Lieb-Schultz-Mattis-Oshikawa-Hastings (LSMOH) theorem \cite{2000PhRvL..84.3370O,Hastings:2003zx} (for more recent developments, see \eg~\cite{Else:2019lft}), and we can call this an LSMOH constraint.  

A perhaps-simpler example is the free fermion topological insulator in $D=3+1$, protected by charge conservation and time-reversal symmetry.  In this case, the bulk effective action governs a single massive Dirac fermion; a boundary is an interface where the mass changes sign, at which a single Dirac cone arises.  A single Dirac cone in $D=2+1$ realizes the so-called parity anomaly.  
The fact that anomaly transcends a phase of matter is illustrated by the fact that, in the presence of interactions or disorder, 
there are other possible edge theories for the topological insulator.

There is by now a sophisticated (still conjectural) mathematical classification of SPTs for various $G$ in various dimensions \cite{Kapustin:2014gma,xiong2018minimalist} 
about which I will not say more here.
My point is that we are still using their realization of symmetries to label these phases!

\subsection{Anomalies of higher-form symmetries}
\label{sec:anomalies-of-higher-form}

Let's return to the example 
from \S\ref{subsec:SF}
of the $(d-1)$-form symmetry that arises in any superfluid phase \cite{Hofman:2018lfz, Delacretaz:2019brr,Else:2021dhh}.
The current can be written as $(\star J)_\mu = \partial_\mu \varphi$.  
However, in the presence of a background gauge field $A$ for the $\gU(1)$ symmetry, 
the gauge-invariant current is instead 
\be (\star J)_\mu = D_\mu \varphi \ee
where $D_\mu \varphi  = \partial_\mu \varphi - q A_\mu $ is the covariant derivative.  
But this current is not conserved:
\be  d \star J = - q F  \label{eq:SF-mixed-anomaly}\ee 
with $F \equiv dA$.
This equation has a simple interpretation: applying an electric field leads to a supercurrent that increases linearly in time.

The symmetry violation in \eqref{eq:SF-mixed-anomaly} is an example of a mixed anomaly between a $0$-form symmetry and a $(d-1)$-form symmetry,
that arises automatically from SSB.
Reference \cite{Delacretaz:2019brr} shows a converse statement: any system with $\gU(1)^{(0)} \times \gU(1)^{(D-2)}$ symmetry
with anomaly \eqref{eq:SF-mixed-anomaly} contains a Goldstone boson in its spectrum.
Since no long-range order is assumed, this is a more general statement than Goldstone's theorem -- it applies even in $D=2$.  

This perspective can be used to demonstrate the existence of equilibrium states with non-dissipating current \cite{Else:2021dhh}.

A direct 1-form generalization of Oshikawa's argument \cite{2000PhRvL..84.3370O} appears in \cite{Kobayashi:2018yuk}.  This is an example of a mixed anomaly between a 1-form symmetry and lattice translation symmetry.  

We should give an example of an anomaly of a higher-form symmetry that does not involve zero-form symmetries.  
An example is provided by the theory of abelian anyons in $D=2+1$, and is best understood by regarding an anomaly as an obstruction to gauging.   Gauging a continuous 1-form symmetry means coupling the conserved current $J^{\mu\nu}$ to a {\it dynamical} two-form gauge field $b_{\mu\nu}$ by a term like $b_{\mu\nu} J^{\mu\nu}$.  
That is, gauging a symmetry means summing over all possible background fields.
In the discrete case, this is the same as summing over the insertions of all possible symmetry operators.  
(In the continuous case, it also requires summing over connections that are not flat.)

Thus, gauging a 1-form symmetry in 2+1 dimensions 
means proliferating the worldlines of the associated anyons \cite{Gaiotto:2014kfa, Hsin:2018vcg}; 
this is `anyon condensation' \cite{Burnell:2017otf}.
But it only makes sense to condense particles with bosonic self-statistics: condensation means essentially that the many-particle wavefunction is a constant, which has bosonic statistics.
Therefore, a subgroup of a 1-form symmetry generated by line operators with nontrivial statistics cannot be gauged.  
We conclude that, in 2+1 dimensions, the 't Hooft anomaly of a 1-form symmetry is encoded in the self-statistics of the line operators, \ie~of the anyons.
Thus, the algebra \eqref{eq:projective-laughlin} is an example of a 1-form symmetry with an 't Hooft anomaly.
Notice that from this point of view, non-trivial mutual statistics of a pair of anyon types $a$ and $b$ is a mixed 't Hooft anomaly: 
it does not stop us from gauging (\ie~condensing) $a$, but we cannot condense both simultaneously, since in the presence of the $a$ condensate, $b$ is confined.  
The algebra for discrete gauge theory \eqref{eq:Z_p-gauge-theory-algebra} can also be regarded an example of
an anomaly for higher-form symmetry because the charged operators $V_n$ are also topological;
so this is a $1$-form symmetry and a $(D-2)$-form symmetry with a mixed anomaly.
In fact, the generalized symmetry that emerges and is spontaneously broken in any topologically ordered groundstate is always anomalous: this is the statement of {\it braiding nondegeneracy}, which is an axiom of topological field theory, and a theorem of Entanglement Bootstrap \cite{Shi:2023kwr}.

\subsection{SPT phases of higher-form symmetries}

We can combine the ingredients of the above discussions, and consider SPT phases protected by higher-form symmetries \cite{Kapustin:2013uxa}.  
This is a slightly awkward subject because higher-form symmetries tend to be emergent, and it therefore might be artificial to restrict ourselves to the subspace of Hamiltonians with exact higher-form symmetry.  

In $D=2+1$, an 't Hooft anomaly for a 1-form symmetry is diagnosed by the self-statistics of the line operators.
So the edge of a 1-form $G$ SPT in $D=3+1$ just needs to have $G$ topological order with 
quasiparticles that aren't bosons.
Lattice models for higher-form SPTs have been written down in 
\cite{PhysRevB.93.155131,Tsui:2019ykk} and 
effective theories studied in 
\cite{Hsin:2021qiy}.

\section{Subsystem symmetries and fracton phases}

Above we have discussed $p$-form symmetries, described by symmetry operators acting on codimension-$(p+1)$ submanifolds of spacetime.  
These operators were {\it deformable}, in the sense that their correlations only depend on their deformation class in spacetime (avoiding any charged operator insertions).

A distinct generalization of the notion of symmetry arises by defining symmetry operators acting independently on {\it rigid} subspaces of the space on which the system is defined.  
That is, we can imagine that there is a different symmetry operator for each subspace, even in the same homology class, 
so that the symmetry operators are not topological, but still commute with the Hamiltonian.
This is sometimes called a ``faithful" symmetry \cite{Qi:2020jrf} or {\it subsystem symmetry}.
This generalization is not compatible with Lorentz invariance.  

An object charged under such a subsystem symmetry cannot leave the locus on which the symmetry is defined.  
This sort of restricted mobility of excitations is a defining property of fracton phases \cite{nandkishore2019fractons,pretko2020fracton}.  
A fracton phase can be identified as one that spontaneously breaks
such a ``faithful" higher-form symmetry
\cite{Qi:2020jrf, Shen:2021rct, Rayhaun:2021ocs}.
Foliated fracton phases \cite{2018PhRvX...8c1051S} like the X-cube model \cite{2016PhRvB..94w5157V}
spontaneously break a `foliated 1-form symmetry' acting independently on each plane  of a lattice \cite{Qi:2020jrf}.

A closely-related concept is that 
of multipole symmetries ({\it e.g.}~\cite{Pretko:2016kxt, Gromov:2018nbv, Seiberg:2020bhn, Seiberg:2020wsg, Seiberg:2020cxy, Stahl:2021sgi}).  A multipole symmetry 
is one where the continuity equation involves extra derivatives, like $ \partial_0 J^0 + \partial_i \partial_j J^{ij} = 0 $ (a dipole symmetry).
Such a conservation law produces conserved charges that need not be integrated over all of space, and act independently of each other.  
(For example \cite{Seiberg:2020bhn}, consider the continuity equation $ \partial_0 J^0 + \partial_x \partial_y J =0$ in $D=2+1$; 
then $ Q_x(x) = \int dy J^0(x,y) $ is conserved for each $x$.)
The simplest example is that conservation of dipole moment implies that charges are immobile \cite{Pretko:2016kxt}.  

Models with such symmetries have been studied for a long time in the condensed matter literature \cite{2002PhRvB..66e4526P}.
Efforts to understand how the rules of ordinary field theory must be relaxed to accommodate 
such systems and their symmetries have been vigorous 
(see \eg~\cite{Slagle:2017wrc, 2018PhRvX...8c1051S, Bulmash:2018lid, Bulmash:2018knk, Seiberg:2019vrp, Seiberg:2020bhn, Seiberg:2020wsg, Seiberg:2020cxy,Lake:2021pdn} and references therein and thereto).
Attempts have been made to classify subsystem-symmetry-protected topological phases
\cite{Devakul:2018fhz} and their anomalies \cite{Burnell:2021reh}, and to understand subsystem-symmetry-enriched topological order
\cite{Stephen:2020zxm}.  A subsystem-symmetry-based understanding of Haah's code \cite{2011PhRvA..83d2330H}
appears in \cite{Williamson:2016jiq}.

An important issue is the robustness of such phases, especially in the gapless case, upon breaking the large symmetry group.   
At least in examples, the scaling dimensions of operators charged under the subsystem symmetry is large, and 
in fact diverges in the continuum limit \cite{2002PhRvB..66e4526P, Seiberg:2019vrp, Seiberg:2020bhn, Seiberg:2020wsg, Seiberg:2020cxy} (see in particular Eq.~(121) of the first reference).  This shows that there is at least a small open set in the space of subsystem-symmetry-breaking couplings in which such phases persist.  

\begin{figure}[h!]
$$\parfig{.4}{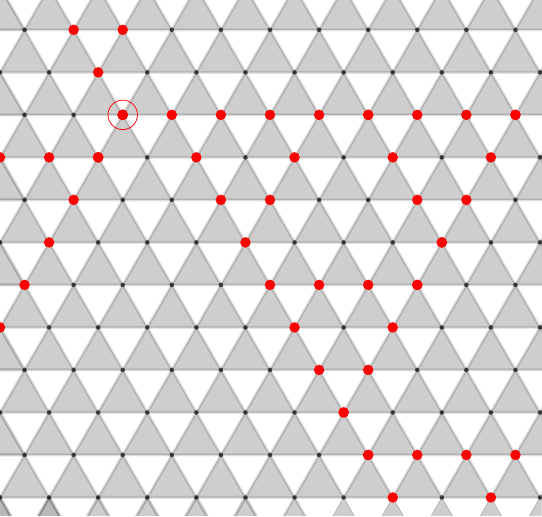}$$
\caption{An example of the support of a fractal symmetry operator in the Newman-Moore model.  
If we flip only the red spins, it preserves the Hamiltonian \eqref{eq:newman-moore}.
That is, every up-triangle has an even number of red dots.  There are many ways to accomplish this.
}
\label{fig:fractal-symmetry}
\end{figure}

{\bf Fractal symmetry.}  The subsystem on which a symmetry acts can be more interesting than just a line or a plane.  For example,
it can be a fractal \cite{Yoshida:2013sqa, devakul2018fractal}.
The Newman-Moore model  \cite{newman1999glassy} is a simple example of a model with a symmetry operator supported on a fractal subset of space. 
Put qubits on the sites $i$ of the triangular lattice and consider 
\be\label{eq:newman-moore}
H = \sum_{ijk \in \Delta} Z_iZ_jZ_k + g \sum_i X_i ,\ee
where the sum is only over up-pointing triangles.
To see that this has a fractal symmetry, pick a spin to flip, say the circled spin in Fig.~\ref{fig:fractal-symmetry}.
Moving outward from that starting point and demanding that each up-triangle contains an even number of flipped spins, 
there are many possible self-similar subsets of the lattice we can choose to flip.
In fact, there is an extensive number.

This transverse-field Newman-Moore model \eqref{eq:newman-moore}  has a number of interesting properties.
It has a self-duality mapping $g \to 1/g$, obtained by defining dual spins $\tilde X_\Delta\equiv \prod_{i \in \Delta} Z_i Z_j Z_k $ on a new lattice with sites corresponding to the up-pointing triangles.  
The exotic critical point at $g=1$\footnote{Earlier work
\cite{Yoshida:2013sqa, vasiloiu2020trajectory} found indications of a first-order transition.} \cite{Zhou:2021wsv} separates a gapped paramagnetic phase from a gapless phase in which the fractal $\IZ_2$ symmetry is spontaneously broken.
Such critical points were claimed \cite{Zhou:2021wsv} to be `beyond renormalization'; rather, what is broken is the connection between short distances and high energies \cite{Lake:2021pdn}.
Other models with such fractal symmetry have been studied in \cite{Myerson-Jain:2021wuh}.

\section{Categorical symmetries}

\label{sec:cat}
Our understanding of what is a symmetry of a quantum many body system or quantum field theory (QFT) has evolved quite a bit.  The above discussion shows that the presence of a symmetry means the existence of {\it topological defect operators}\footnote{A sufficient condition for this conclusion is Lorentz symmetry.  In its absence, we have already seen examples of systems with subsystem symmetries,  where there are  
operators that commute with the Hamiltonian that are not fully topological.}.  (I believe the word `defect' in this name just refers to the fact that these operators have positive codimension.)
In the case of an ordinary symmetry, these are the symmetry operators, $U_g(\Sigma_d)$, that we discussed above; they are 
labelled by a group element $g\in G$, and supported on a codimension-one (\eg~fixed-time) slice $\Sigma_d$, and are topological in the sense that their correlation functions do not change under continuous deformations.  
These operators satisfy a `fusion rule' in the sense that for two symmetry operators associated with the same time-slice,
 $\lim_{\eps \to 0^+} U_g(t+ \epsilon) U_h(t) = U_{gh}(t) $.  
When a local operator crosses such a $U_g$, it gets acted on by the transformation $g$.

If the surface $\Sigma_d$ is not a fixed-time slice, such an operator implements a modification of the Hamiltonian, such as a change of boundary conditions.A good example to keep in mind is 
the defect operator $U_{-1}(\Sigma)$ in the classical Ising model.  
It is an instruction to flip the sign of the coupling along any bond crossing the codimension-one locus $\Sigma$.  
This operator is topological: deforming $\Sigma$ through a region $R$ is accomplished by redefining all the spins in $R$ by $\sigma \to -\sigma$.  
This shows that the charged operator is the spin.

A useful perspective is to reverse the logic, and regard the existence of topological defect operators 
as the definition of a symmetry.  One of the advantages of this perspective is that it treats continuous and discrete symmetries uniformly;
it also makes no reference to transformations of fields, and so treats Noether symmetries and topological symmetries uniformly.
And from this perspective it is easy to see some generalizations.  
The first generalization is that a $p$-form symmetry is associated with (unitary) topological operators whose support has codimension $p+1$.  
In the low-energy theory describing an abelian topological order in $D=2+1$, these operators $U$ are the holonomies of anyon worldlines.
For two operators on the same submanifold, 
$U_\alpha(M_{D-p-1})U_\beta(M_{D-p-1}) = U_{\alpha + \beta}(M_{D-p-1})$
where for $p>1$, the order does not matter, and we adopt an additive notation.

One way in which an anomaly can appear in this language is when, in the presence of background fields $F$ (which could include curvature of spacetime), 
the symmetry operator fails to be topological in the sense that 
\be U_\alpha(M) = e^{ \ii  \alpha \int_{D} \Gamma(F) } U_\alpha(M'),\ee
where $D$ is a $(D-p)$-dimensional surface bounding the difference of $M$ and $M'$: $ \partial D = M- M'$,
and $\Gamma(F)$ is some polynomial in background fluxes and curvatures of the appropriate form degree \cite{Chang:2018iay}.

The preceding discussion suggests a further generalization, which we will need in order to describe non-abelian topological order as SSB: what about the worldlines of non-abelian anyons?  
This is a dramatic step because the algebra of topological operators $T_a$ that transport non-abelian anyons is no longer a group.  Rather, they satisfy the fusion algebra:
\be T_a T_b = \sum_c N_{ab}^c T_c . \label{eq:fusion-algebra}\ee
By definition, a topological order is non-abelian if there is more than one term on the RHS of this equation for some choice of $a,b$.  
Whereas multiplication of two elements of a group always produces a unique third element, here we produce a superposition of elements, weighted by {\it fusion multiplicities} $N_{ab}^c$.  
Further,  there is some tension between the fusion algebra \eqref{eq:fusion-algebra}
and unitarity of the operators $T_c$.  The trivial anyon corresponds to the identity operator, $T_1 = 1$.  
Each type of anyon $a$ has an antiparticle ${\bar a}$.
Since $T_{\bar a}$ corresponds to transporting $a$ in the opposite direction, we expect that $T_{\bar a} = T_a^\dagger$, and therefore
\eqref{eq:fusion-algebra} says in particular
\be T_a T_a^\dagger = \sum_c N_{a \bar a}^c T_c .\ee
If the RHS here has a term other than $N_{a\bar a}^1$, then $T_a$ is not unitary.  
As an example, consider the  Ising topological order, with three anyon types $\{1, \psi, \sigma \}$ and 
the fusion rules 
\be \label{eq:ising} T_\sigma T_\sigma = 1 + T_\psi,  ~~~T_\sigma T_\psi = T_\psi T_\sigma = T_\sigma, ~~~T_\psi T_\psi = T_1 .\ee
Note that $\sigma$ is its own antiparticle.  
\eqref{eq:ising} implies that the topological line operator $T_\sigma$ cannot be unitary, and moreover cannot be inverted by any linear combination of $T_a$.  
Such symmetries are called {\it categorical symmetries} or {\it fusion category symmetries}.  

More generally, any algebra of topological operators acting on a physical system can be regarded as encoding some kind of generalized symmetry.

At the moment, condensed matter applications of the idea of fusion category symmetries remain in the realm of relatively formal developments,
as opposed to active phenomenology of real materials. 
One application is to understand non-abelian topological order as spontaneous symmetry breaking\footnote{A related perspective appears via the `pulling-though' operators 
in the tensor network description of topological orders reviewed in \cite{Cirac:2020obd}.  For a study of categorical symmetries realized as matrix product operators, see \cite{Garre-Rubio:2022uum}.}.
A concrete example of a (2+1)d model with non-invertible symmetries is $G_k$ Chern-Simons (CS) theory, with non-Abelian gauge group $G$ at level $k>1$.  
The non-invertible symmetry operators are the Wegner-Wilson lines.
The specific example of $\gSU(2)_2$ CS theory can describe the Ising topological order, and is possibly realized as part of the effective low-energy description of $\nu = {5 \over 2 }$ quantum Hall states.

More generally, any topological field theory for non-Abelian topological order 
enjoys such a non-invertible symmetry.  
A nice example of the application of this perspective on anyon worldlines as symmetry operators 
is \cite{Kaidi:2021gbs} which provides a 
condition on the anyon data 
required for a general 2+1D topological order to admit a gapped boundary condition, beyond vanishing chiral central charge.

\begin{figure}
$$
a) \parbox{.2\textwidth}{
\begin{tikzpicture}[auto, node distance=.25cm]
	\node (A) at (1,1) {};
	\node (B) [above right of = A]{$\beta$};
	\node (A) at (-1,1) {};
	\node (B) [above left of = A]{$\alpha$};
	\node (A) at (0,-1) {};
	\node (B) [below of = A]{$\gamma$};
	\draw[very thick, red] (0,0) -- (1,1); 
	\draw[very thick,blue] (0,0) -- (-1,1); 
	\draw[very thick, purple] (0,-1) -- (0,0); 
\end{tikzpicture}
}
~~~~~~~~~~~~ b)
\parbox{.2\textwidth}{
\begin{tikzpicture}[auto, node distance=.25cm]
	\node (A) at (1,1) {};
	\node (B) [above right of = A]{$\gamma$};
	\node (A) at (-1,1) {};
	\node (B) [above left of = A]{$\alpha$};
	\node (A) at (0,-1) {};
	\node (B) [below of = A]{$\delta$};
	\node (A) at (0,1) {};
	\node (B) [above left of = A]{$\beta$};
	\node (A) at (.25,.25) {};
	\node (B) [above left of = A]{$\mu$};

	\draw[very thick] (0,0) -- (1,1); 
	\draw[very thick] (0,0) -- (-1,1); 
	\draw[very thick] (0,-1) -- (0,0); 
	\draw[very thick] (.5,.5) -- (0, 1);
\end{tikzpicture}
}
= 
\sum_\nu \( F^{\alpha\beta\gamma}_\delta \)_{\mu\nu}
\parbox{.2\textwidth}{
\begin{tikzpicture}[auto, node distance=.25cm]
	\node (A) at (1,1) {};
	\node (B) [above right of = A]{$\gamma$};
	\node (A) at (-1,1) {};
	\node (B) [above left of = A]{$\alpha$};
	\node (A) at (0,-1) {};
	\node (B) [below of = A]{$\delta$};
	\node (A) at (0,1) {};
	\node (B) [above right of = A]{$\beta$};
	\node (A) at (-.25,.25) {};
	\node (B) [above right of = A]{$\nu$};

	\draw[very thick] (0,0) -- (1,1); 
	\draw[very thick] (0,0) -- (-1,1); 
	\draw[very thick] (0,-1) -- (0,0); 
	\draw[very thick] (-.5,.5) -- (0, 1);
\end{tikzpicture}
}
$$
\caption{a) Fusion of symmetry operators: this junction is allowed if $N_{\alpha\beta}^\gamma \neq 0$.  b) Associativity data of fusion of symmetry operators (in the simpler case where the fusion coefficients $N_{\alpha\beta}^\gamma$ are only $0$ or $1$).}
\label{fig:branched}
\end{figure}

Part of the reason for the nomenclature `categorical symmetry' is that such a collection of symmetry operators comes with some additional data.
Besides putting two symmetry operators right on top of each other, we can also consider symmetry operators associated with branched manifolds, as in Fig.~\ref{fig:branched}a.  Once we allow such objects, we must also consider more complicated objects related to the associativity of the product, 
as in Fig.~\ref{fig:branched}b, which relates the two ways of resolving a 4-valent junction of topological operators into two 3-valent junctions.  This associativity information (creatively called $F$-symbols) is part of the specification of the categorical symmetry, 
and must satisfy the pentagon identities (see {\it e.g.}~Fig.~1 of \cite{Thorngren:2019iar}).  In the case of 1-form symmetry in (2+1)-D, there is further information associated with braiding.

\begin{figure}
$$
\begin{tikzpicture}[auto, node distance=.35cm]
	 \draw[very thick] (0,-1) -- (0,1); 
	 \draw[fill=black] (-1,0) circle (0.05); 
	 \node (A) at (0,-.8) {};
	 \node (B) [left of = A]{$\CN$};
	 \node (A) at (-1,0) {};
	 \node (B) [above right of = A]{$\sigma(x)$};

\node at (1,0) {$ \rightsquigarrow $};

	\begin{scope}[xshift=2cm]
		 \draw[very thick] (0,-1) -- (0,1); 
		 \draw[fill=black] (1,0) circle (0.05); 
		 \draw[dashed] (0,0) -- (1,0); 
		 \node (A) at (0,-.8) {};
		 \node (B) [left of = A]{$\CN$};
		 \node (A) at (0,.8) {};
		 \node (B) [left of = A]{$\CN$};
		 \node (A) at (1,0) {};
		 \node (B) [above right of = A]{$\mu(x)$};
		 \node (A) at (.5,0) {};
		 \node (B) [below of = A]{$\eta$};
	\end{scope}
	 
\end{tikzpicture}
$$
\caption{When a spin $\sigma$ moves through the duality wall $\CN$, it turns into a disorder operator $\mu$, attached by a topological line $\eta$ to the duality wall.  The right figure illustrates the fact that $N_{\CN\CN}^\eta \neq 0$.}
\label{fig:duality-wall}
\end{figure}

A good example of a non-invertible line operator appears in the critical Ising conformal field theory (CFT) in $D=1+1$, in the form of the {\it duality wall} (Fig.~\ref{fig:duality-wall}).  
The definition of such an object is: when we pass through the wall, we act by the Kramers-Wannier self-duality interchanging the spin and the disorder operator.  The latter is not a local operator, but rather must be attached to a branch cut across which the $\IZ_2$ symmetry acts.  
Moving a local spin operator through such a duality wall then turns it into an operator attached by a topological defect line to the duality wall.  
The fusion algebra of the duality wall operator $\CN$ and the ordinary $\IZ_2$ symmetry line operator $\eta$ can be summarized as 
$$ \eta \eta = 1, ~~ \CN \eta  = \eta \CN = \CN, ~~\CN \CN = 1 + \eta .$$
(These are a relabelling of the Ising fusion rules above.)
The last, non-abelian, relation comes from the fact that the Kramers-Wannier duality only keeps track of the locations of domain walls, and erases the information about the overall spin flip.  
In a theory with such a symmetry operator, 
RG flows generated by a perturbation by a local operator can only generate operators that pass freely through the wall \cite{Chang:2018iay, Komargodski:2020mxz}.  
 Examples of duality walls in $D=3+1$ were studied in 
\cite{Choi:2021kmx, Kaidi:2021xfk}.

Categorical symmetries have been studied most extensively in 1+1d QFTs (\eg~\cite{Bhardwaj:2017xup, Chang:2018iay, Thorngren:2019iar, Lin:2019hks,Gaiotto:2020iye,
Komargodski:2020mxz, Thorngren:2021yso,Kikuchi:2021qxz,Burbano:2021loy}), where they can be used constrain RG flows.  
It was shown in 
\cite{Chang:2018iay}
that certain non-invertible symmetries can forbid a trivial gapped groundstate,
as in the LSMOH theorem.
The idea is to consider the partition function on $T^2$ with a symmetry line operator $\CL$ wrapping one of the circles, and argue by contradiction.  
If there is a gap, we can evaluate this quantity in the effective low-energy topological theory.
Demanding modular invariance (\ie~that we get the same answer whichever circle we regard as time) relates 
the trace over the Hilbert space with twisted boundary conditions 
\be 
\tr_{{\cal H}_{{\cal L}}} e^{ - \beta (H-E_0)} 
=
\parbox{.085\textwidth}{\begin{tikzpicture}
	\filldraw[fill opacity = .1] (0,0) rectangle (1,1); 
	\draw[->] (.4999,0) -- (.5,0) ; 
	\draw[->] (.4999,1) -- (.5,1) ; 	
	\draw[->] (0,.45) -- (0,.45) ; 
	\draw[->] (1,.45) -- (1,.45) ; 	
	\draw[->] (0,.55) -- (0,.55) ; 
	\draw[->] (1,.55) -- (1,.55) ; 	
	\draw[very thick, darkblue] (0,.65) -- (1,.65); 
	\node () at (.2, .15) {$t$}; 
	\node () at (.85, .25) {$x$}; 
	
	\node[darkblue] () at (.5, .82) {${\cal L}$};
	
\end{tikzpicture}
}
= 	
\parbox{.085\textwidth}{
\begin{tikzpicture}
		\filldraw[fill opacity = .1] (0,0) rectangle (1,1); 
	\draw[->] (.4999,0) -- (.5,0) ; 
	\draw[->] (.4999,1) -- (.5,1) ; 	
	\draw[->] (0,.45) -- (0,.45) ; 
	\draw[->] (1,.45) -- (1,.45) ; 	
	\draw[->] (0,.55) -- (0,.55) ; 
	\draw[->] (1,.55) -- (1,.55) ; 	
	\draw[very thick, darkblue] (0,.65) -- (1,.65); 
	\node () at (.2, .15) {$x$}; 
	\node () at (.85, .25) {$t$}; 
	\node[darkblue] () at (.5, .82) {${\cal L}$};
\end{tikzpicture}
}
= 
\tr \CL  e^{ - {2\pi \over \beta} (H-E_0)},
\ee
to the ordinary trace with the insertion of the symmetry operator.  
In a topological field theory, the former quantity is just $ \tr_{\CL} 1$, the number of states in the twisted sector.  
If there is furthermore a unique groundstate, then the latter quantity is just $ \vev{\CL }$.
Since the former is a non-negative integer, we can conclude that if $ \vev{\CL} $ is not a non-negative integer, then there cannot be a unique gapped groundstate.  For example, a certain perturbation of the tricritical Ising model has a symmetry  operator $W$ with (Fibonacci) fusion algebra $W^2 = 1+ W$.  This algebra implies that the eigenvalues of $W$ are $ (1 \pm \sqrt{5})/2 $, and there must therefore be an even number of vacua for its expectation value to be an integer.  
More generally, this argument shows that (in a system with modular invariance) the existence of a symmetry operator with no integer eigenvalues forbids a unique groundstate \cite{Chang:2018iay}:
if the unique groundstate were not an eigenvector of $\CL$, we could make another groundstate by acting with $\CL$, therefore if there is a unique groundstate, then $ \vev{\CL}$ must be an eigenvector of $\CL$.  
A related argument shows \cite{Pal:2020wwd} that all irreps of $G$ appear in the spectrum of a 1+1d CFT with finite symmetry group $G$.
An extension of this modular-invariance argument to 3+1-D can be found in \cite{Choi:2021kmx}.

The edge theory of the $G_k$ CS theory is the $G_k$ WZW model; it inherits the categorical symmetry from the Wegner-Wilson lines running parallel to the boundary.
These ingredients are used by \cite{Komargodski:2020mxz} to construct massless 2d QCD with adjoint fermions by coupling CS theory on an interval to 2d Yang-Mills theory; the construction makes manifest some surprising non-invertible symmetries of the theory, which guarantee deconfinement.

\cite{Thorngren:2019iar,Gaiotto:2020iye} argue that a 1+1d system with fusion category symmetry can always be realized as a boundary condition of a 
gapped 2+1 dimensional topological order with anyon types carrying the associated labels of the topological operators.
They wish to study {\it anomalies} of the fusion category symmetry, to use them as RG invariants, and to label SPTs protected by such a symmetry:
the bulk is a realization of anomaly inflow.
Gapped edge theories are realized if the bulk theory admits gapped boundary conditions; 
such a bulk theory has an exactly solvable description as a string-net model \cite{Levin:2004mi}.  
Explicit lattice models for gapped phases in $D=1+1$ with fusion category symmetries appear in \cite{Inamura:2021szw}.

Examples of systems with categorical symmetries include the anyon chain models studied in \cite{Feiguin:2006ydp}, which uses 
the categorical symmetry to explain the gaplessness of the model.  
\cite{Aasen:2016dop,Aasen:2020jwb, Vanhove:2021zop}
build classical lattice models whose defects realize a fusion category.

The terms `categorical symmetry' and `non-invertible symmetry' are not used in a unique way in the literature.
In \cite{Ji:2019jhk, 2020PhRvR...2d3086K,2021PhRvR...3c3024Z}, the term is used in the context of gapped phases in $D=2+1$ with gapless boundaries; 
the idea is that such edge theories can have anomalies that go beyond those associated with invertible phases, which are therefore called non-invertible anomalies.   
The term `algebraic higher symmetry' is used in 
\cite{Ji:2019jhk, 2020PhRvR...2d3086K} for the concept I called categorical symmetry above.
\cite{Ji:2019jhk, 2020PhRvR...2d3086K, Chatterjee:2022kxb} argue that the most general notion of symmetry of a $D$-dimensional system 
is labelled by a topological order in one higher dimension.

\section{Gapless states}

\subsection{Critical points}

The second part of the Landau paradigm (Item \ref{item2})
says that at a critical point, the critical degrees of freedom are the fluctuations of an order parameter.  
Apparent exceptions to this statement come in several varieties.  

First, any transition out of a phase without a local order parameter presents an immediate problem.
Consider the case of $\IZ_2$ gauge theory in $D=2+1$, which spontaneously breaks a $\IZ_2$ 1-form symmetry,
with a charged loop operator $W[C]$.  Can we understand the critical theory in terms such a string order parameter field?
By Wegner's duality \cite{Wegner1971}, the local physics of the critical theory is in the same universality class as the 3d Ising model.  
This is yet another point of view from which the 3d Ising model should have a string theory dual
\cite{Polyakov:1987ez, Iqbal:2020msy}.

Second, there are direct transitions between states that break different symmetries, known as deconfined quantum critical points (DQCP) (\cite{Wang:2017txt} 
has a useful summary and references).  Does this require a revision to Item \ref{item2} of the Landau Paradigm as stated above?
There is a sense in which the degrees of freedom of the critical theory 
{\it are} simply the order parameters of both of the neighboring phases, coupled by a WZW term \cite{tanaka2005many,senthil2006competing}.
The presence of the WZW term is required by a mixed anomaly between the two symmetries.  
It says that defects of the order in one phase carry charge under the other \cite{senthil-levin}.
This perspective predicts a dramatic enlargement of symmetry at the critical point, not obvious from other points of view, and borne out by numerical work.
This symmetry-based description as a non-linear sigma model has the serious shortcoming that it is strongly coupled, 
but so is the more-familiar description in terms of abelian gauge theory.

Independently of the extended Landau Paradigm, I should also mention that the study of order parameters for higher-form symmetry at various critical points has been instructive
\cite{2021PhRvR...3c3024Z,2021PhRvR...3c3024Z,Wang:2021lmb,Wang:2021yaf}.  In particular, their study has provided independent evidence 
that the 2+1d DQCP between Neel and VBS phases is a weakly first order transition \cite{Wang:2021yaf}.

\subsection{Gapless phases}

Gapless phases are a wild frontier of our understanding, and we certainly do not have a symmetry-based (or any other) understanding of all possibilities at the moment.  
I limit myself to remarks on two illuminating examples.   

First, I mention a set of exotic gapless fractonic phases that can be constructed by assembling layers of quantum Hall states.  
They can be described at low energies by an abelian Chern-Simons theory 
$ S[a^I] = { K^{IJ} \over 4 \pi} \int  a^I \wedge d a^J $
with a nearly-diagonal $K$ matrix whose size grows with the number of layers \cite{qiu1989phases, 2000PhRvL..85.5408N, Ma:2020svo}.  For some choices of $K$ matrix, this represents a new class of gapped fracton phase, with irrational particle statistics and a large-order fusion group.  
For other choices of $K$ matrix, the spectrum is gapless.  
Ref~\cite{Sullivan:2021rbk} shows that the gapless examples of such states can be understood in terms of {\it weak symmetry breaking} \cite{2013PhRvB..88x5136W}.  
This means that the charged operator that condenses is not local, but rather an extended operator, in this case extended along the direction of the stack of layers.

Second, among the list above of apparent exceptions to the Landau Paradigm, it remains to discuss the Landau Fermi Liquid.  
\cite{Else:2020jln} gives something like a symmetry-based understanding of both Fermi liquids and a large class of non-Fermi liquids (for a review of the latter, see \cite{2018ARCMP...9..227L}).  
First we assume translation symmetry, so that we may speak about a well-defined Fermi surface in momentum space.  
The key ingredient is an emergent symmetry representing independent particle number conservation at each point on the Fermi surface.  
In 2+1 dimensions, where the simplest Fermi surface is a circle, this is a {\it loop group} symmetry; that is, the symmetry transformation is a map from the circle to $\gU(1)$.  
Such a loop-group symmetry emerges in the Landau theory, as well as in a large class of non-Fermi liquids obtained by coupling a Fermi surface to gapless modes.  
\cite{Else:2020jln} shows that a state with a fractional and continuously-variable filling {\it must} have such a large symmetry.
From this starting point, the authors develop an understanding of Luttinger's theorem as an anomaly of this loop symmetry.
(A related anomaly-based perspective on Luttinger's theorem appears in \cite{Ma:2021isw}.)
It shows that in a system with such a loop group symmetry, a literal Fermi arc, \ie~a boundary of the Fermi surface, would imply a violation of charge conservation: the Fermi surface must be the boundary of some region of the Brillouin zone.

\section{Concluding remarks}

{\bf Topological local operators.}  What about the case of $(D-1)$-form symmetries in $D$ spacetime dimensions? 
This means that there are {\it local} operators that are topological. 
This case is studied in \cite{Pantev:2005rh, Hellerman:2006zs, Sharpe:2015mja} and more generally in \cite{Komargodski:2020mxz, Delmastro:2021otj, Cherman:2021nox}.  
The conclusion is that the Hilbert space of such a system is divided into superselection sectors with different values of the topological operators.
An example where this arises is in gauge theory in $D=1+1$ without minimally-charged matter, where sectors represent different values of the electric flux.
\cite{Cherman:2021nox} considers what happens when the action is perturbed by such operators, which are always relevant.
The perturbation changes the difference of the vacuum energies between different sectors.

{\bf Higher groups.}   
The concept of higher groups can be regarded as a natural extension of higher-form symmetry (see \eg~\cite{2003math......7200B} for a broader mathematical perspective).
For example, a $2$-group structure can be defined in a physical context as follows: it is a modification of the current algebra of a $1$-form symmetry and a $0$-form symmetry, so that 
the $0$-form gauge transformation acts nontrivially on the $2$-form background field $B$ for the $1$-form symmetry:
\be\label{eq:2-group} A \to A + d\lambda, ~ B \to B + \kappa \lambda dA ,\ee
where $A$ is the background $1$-form field for the $0$-form symmetry, and 
$\kappa$ can be regarded as a structure constant.  
This construction is closely related to the Green-Schwarz mechanism of anomaly cancellation:
suppose, for example, the effective action of a $(1+1)$D theory with the above ingredients 
has an anomalous variation $\delta_\lambda S =  \int \kappa \lambda {dA\over 2\pi} $ under a $0$-form gauge transformation.
Then the modified action $  S- \int {B \over 2 \pi} $ is invariant under the transformation \eqref{eq:2-group}.
Though it has not yet explicitly played a role in the condensed matter literature to my knowledge, 
it appears in many places in QFT 
(\eg~\cite{Cordova:2018cvg,Benini:2018reh,Bhardwaj:2017xup,Iqbal:2020lrt})
and we can expect that it will be useful.

{\bf Other applications.}  In the preceding discussion, we have focussed on generalizations of notions of symmetry as applied to zero-temperature groundstates of quantum matter.  
I should mention that these same generalized symmetries have a number of other applications:
\begin{itemize}
\item A new organizing principle for magnetohydrodynamics \cite{Grozdanov:2016tdf, Armas:2018zbe,Armas:2018atq,Grozdanov:2017kyl}.
More generally, many kinds of exotic hydrodynamics can be understood by applying the systematic logic of hydrodynamics 
to a system with generalized symmetries (see, for example, \cite{Grozdanov:2018ewh}).

\item \cite{Sogabe:2019gif} provides a nice example using both anomalies and generalized symmetries to understand the spectrum of Goldstone modes of the Standard Model in a magnetic field, and suggests a realization of the same physics in Dirac semimetals.

\item More generally, more symmetry means more possible anomalies, and therefore 
new anomaly constraints on IR behavior of QFT. 
For example, a mixed anomaly between time-reversal symmetry and a 1-form symmetry implies an LSMOH constraint on 
the groundstate of Yang-Mills theory at $\theta = \pi$ \cite{Gaiotto:2017yup, Wan:2018zql, Wan:2019oyr, Cordova:2019bsd}.
Work using anomalies involving higher-form symmetries to constrain dynamics of QFT includes \cite{Gaiotto:2017tne,Kitano:2017jng,Tanizaki:2017mtm,Komargodski:2017dmc,Anber:2018iof,Wan:2018djl, Cordova:2018acb, Cherman:2019hbq,Cox:2021vsa,Nguyen:2021naa} and many others.

\end{itemize}

{\bf Disorder.}  I have not spoken about systems with disorder.  Even if we are generally interested in clean systems,
it is important to ask about the stability of our statements to the introduction of disorder.  
In the case of zero-form symmetries, the Imry-Ma argument for stability of SSB proceeds by coupling 
the local order parameter to the disorder.  
Naively, the inability to write such a coupling corroborates our expectation that higher-form SSB is even more robust
\cite{Stahl:2021sgi}.

{\bf Dynamics.} I have focused entirely on equilibrium phases of matter.  Dynamics of quantum matter is a current frontier, in which of course symmetries continue to play a crucial role.  
A generalization of the notion of symmetry that has appeared in this context 
is the phenomenon of Hilbert space fragmentation: 
this is what happens when the algebra of operators that commute with each term of the Hamiltonian 
grows exponentially with system size \cite{2021arXiv210810324M} (for systems with ordinary symmetries, this algebra grows only polynomially with system size).

{\bf Still beyond Landau?}
In this review, I've tried to motivate the following question:
Does the enlarged
Landau paradigm (including all generalizations of  symmetries, and their anomalies)
incorporate all equilibrium quantum phases of matter 
(and transitions between them)
as consequences of symmetry? 
Even if the answer is `no', I think it has already been a fruitful question.
I close by enumerating some outstanding possible exceptions to even the most generous interpretation
in hopes of encouraging some further thought in this direction.

\begin{itemize}
\item 
Symmetries can forbid all relevant operators that would lift gapless modes that are however not Goldstones.
An example is chiral symmetry in QCD, which forbids fermion masses.
A condensed matter example is the Dirac spin liquid -- a phase described by a CFT with no symmetric relevant operators.

\item 
Above I argued that the DQCP between two distinct symmetry-breaking phases satisfies Item \ref{item2} of the Landau Paradigm
because it admits a description in terms of a nonlinear sigma model whose fields are the order parameters of the two phases.
Ref.~\cite{Zou:2021dwv} generalizes this description to a sigma model on the Stiefel manifold, the coset space $ \gSO(N+4)/\gSO(4)$.
For $N=1$ this is the DQCP, for $N=2$ they give evidence that this is a description of a Dirac spin liquid in terms of only gauge-invariant variables.
The case $N=2$ is called Stiefel liquid, and \cite{Zou:2021dwv} provides a candidate microscopic realization
and argues that it has no weakly-coupled limit.

\item 
An extremely interesting example of a claimed exception to Item \ref{item2} of even the Generalized Landau Paradigm 
is provided by phase transitions described by IR-free gauge theory
\cite{Bi:2019ers}.  
The claim of \cite{Bi:2019ers} is that $\gSU(N)$ gauge theory with adjoint fermions (take $N=2$) 
has a $\IZ_2$ symmetry, and 
describes, as the fermion mass changes sign, a completely novel critical theory for the transition from the trivial phase to the ordinary SSB phase. 
The degrees of freedom of this theory certainly go beyond the fluctuations of the order parameter.   
Notice that for any nonzero mass there is an extra emergent 1-form symmetry associated with the center of the gauge group.
A physical consequence of this symmetry (and a mixed anomaly), were it exact, would be that a domain wall between the two $\IZ_2$-breaking vacua would satisfy an LSMOH constraint:
that is, the domain walls of the ordered phase would carry some extra degrees of freedom, and this would distinguish this phase from the ordinary SSB phase.  
This symmetry is, however, explicitly broken by the massive charged matter of the gauge theory.

\end{itemize}

\vskip 5mm
\section*{Acknowledgements}
I am deeply grateful to Nabil Iqbal for our collaboration, which has had a decisive influence on the perspective advocated by this article.
I would also like to thank Tarun Grover, Diego Hofman, Jin-Long Huang,  Zohar Nussinov, Mike Ogilvie,  Gerardo Ortiz, Leo Radzihovsky, T.~Senthil, Shu-Heng Shao, Zhengdi Sun and David Tong
for conversations about the ideas in this review, and Xiang Li, Dachuan Lu, and Yi-Zhuang You for helpful comments on the manuscript.
This work was supported in part by
funds provided by the U.S. Department of Energy
(D.O.E.) under cooperative research agreement 
DE-SC0009919, and by the Simons Collaboration on Ultra-Quantum Matter, which is a grant from the Simons Foundation (652264).

\appendix
\renewcommand{\theequation}
{\Alph{section}.\arabic{equation}}

\section{Spontaneous symmetry breaking and long-range order}

\label{App:SSB-LRO}

$\ket{\psi}$ is not stationary under the symmetry (SSB)
 if and only if 
there exists   
a charged operator $O$
with $\bra{\psi} O \ket{\psi} \neq 0 $ (long-range order).

Proof:

$\boxed{\Leftarrow}$  
Suppose the state is stationary under the symmetry, meaning
\be S \ket{\psi} = e^{ \ii \alpha} \ket{\psi}. \ee
Then for any charged operator $ O = e^{ \ii \gamma} S^\dagger O S $, $ \gamma \notin 2\pi \IZ$, 
\be  \bra{\psi} O \ket{\psi}  = e^{ \ii \gamma} \bra{\psi} S^\dagger O S \ket{\psi} 
= e^{ \ii \gamma}   \bra{\psi} O \ket{\psi}~,\ee
which says $ \bra\psi O \ket\psi = 0 $, there is no long-range order.

$\boxed{\Rightarrow}$\footnote{I learned this argument from Tarun Grover.}
Consider the reduced density matrix of a region $X$ in the g	roundstate:
\be \rho_X \equiv \tr_{\bar X} \ketbra{\psi}{\psi} = \sum_I \vev{O_I} O_I ,\label{eq:expand-rho}\ee
Here $\{O_I\}$ is a basis of Hermitian operators on $X$, which we can choose to be orthonormal 
with respect to the Hilbert-Schmidt inner product $ \tr O_I O_J = \delta_{IJ}$.
If no charged operator has an expectation value, then the sum in \eqref{eq:expand-rho} only contains neutral operators. 
But then $S \rho_X S^\dagger = \rho_X$, so the state is invariant.  

\thump

Notice that this argument makes no reference to the support of the symmetry operator or its invertibility, and so works also for generalized symmetries.

\section{SSB of higher-form symmetry without topological order, a confession}

\label{App:chamon-state}

In this bonus appendix (added in v3), I want to explain a counterexample \cite{castelnovo2007quantum} to the statement that  SSB of a discrete one-form symmetry implies topological order, based very closely on \cite{Huxford:2023bhb}.
The key point is that SSB of one-form symmetry
\be \ket{\gs_1} = W_C \ket{\gs_2} \label{eq:1SSB}\ee
(or $W_C$ has a perimeter law, as we saw in the preceding appendix)
 is not quite a sufficient condition for topological order.
 That is, \eqref{eq:1SSB} with $W_C$ topological 
 says that the second groundstate {\it can be} obtained from the first by the action of an extended operator; 
but this is not enough to guarantee that there isn't {\it also} some local operator that relates them!

Here is a model, a deformation of the toric code, that provides a counterexample to many simple and nice statements.
It is due to Chamon and Castelnovo \cite{castelnovo2007quantum}.
The Hilbert space is qubits on the links of an arbitrary cell complex, which let's take to be the square lattice for definiteness.  The Hamiltonian is 
\be H_\beta = + \sum_{\text{vertices } i} Q_i - \sum_{\text{plaquettes } p} B_p \ee
where $B_p  \equiv \prod_{\ell \in \partial p} Z_\ell $ is the usual toric code plaquette term and 
\be Q_i \equiv e^{ - \beta \sum_{\ell \in v(i) } Z_\ell} - \sum_{\ell \in v(i) } X_\ell \ee
is a deformation of the star term that depends on a real parameter $\beta$.  
For $\beta \to 0$ this reduces to the usual toric code Hamiltonian (up to an additive constant).

Here are some facts about this model.  
\begin{itemize}
\item The model has a phase transition at $\beta = \beta_c = \half \ln\( 1 + \sqrt{2} \) $.  
The TEE goes from $\log 2$ for $\beta < \beta_c$ to zero for $\beta > \beta_c$.  
Thus, the phase at large $\beta$ is not topologically ordered\footnote{Further evidence for this statement is the fact that the operator that creates a pair of $e$ particles 
$W(C) = \prod_{\ell \in C} Z_\ell $, for an open curve $C$, 
has a nonzero expectation value for $\beta > \beta_c$.  
This expectation value can be mapped to a correlation function in the Ising model 
between two spins at the endpoints of $C$, which becomes long-ranged for $\beta > \beta_c$.  
}.

\item For any groundstate of the toric code $\ket{\gs(0)}$, 
\be \ket{\gs(\beta)} \propto \prod_\ell e^{ \beta Z_\ell/2 } \ket{\gs(0)} \ee
is a groundstate of $H_\beta$.  
This means that on the torus, there is a fourfold degeneracy for every real $\beta$.  

\item In fact the system orders magnetically for $\beta > \beta_c$.
For one thing, the magnetic susceptibility $\partial_\beta\vev{\sum_\ell Z_\ell} $ diverges at the transition.
Further, there is magnetic ordering in 
the sense that the different groundstates $\ket{\gs(\beta)_{ab}}$, 
that come from the eigenstates of $W(C_{x,y})  = \prod_{\ell \in C_{x,y}} Z_\ell $ for various winding
actually have expectation values of the magnetization $\sum_\ell Z_\ell$ that differ by amounts of order $L$.
You can see this from the fact that the wavefunctions weight contributions with different numbers of up spins differently.  

Thus, the different groundstates are actually distinguishable by local operators!

\item The electric one-form symmetry operators $W_C = \prod_{\ell \in C} Z_\ell $ 
(for $C$ a closed curve) still commute with $H_\beta$
(while the magnetic ones do not).  
For every $\beta$, there is SSB of this one-form symmetry in the sense that 
the charged operators have a perimeter law
\be  \bra{\gs(\beta)} V_{\hat C} \ket{\gs(\beta)} \sim e^{ - \beta \ell(\hat C)}  \ee
where $\ell(\hat C)$ is the length of the curve $\hat C$.
(In fact $W_{C}$ also satisfies a perimeter law, but $V_{\hat C}$ is not a symmetry of $H_\beta$, so this is not spontaneously breaking any symmetry.)

\end{itemize}

To see what is happening here, consider the following two requirements of topological order \cite{Huxford:2023bhb}.
For all local operators $\CO$, and candidate orthonormal topological groundstates, 
\begin{enumerate}
\item 
$ \bra{\gs_1}\CO \ket{\gs_1} = \bra{\gs_2} \CO\ket{\gs_2}$.

\item 
$ \bra{\gs_1}\CO \ket{\gs_2} =0$.  

\end{enumerate}

To see that these are distinct conditions, consider 
the states $ \ket{\Uparrow} \equiv \prod_x \ket{\up}_x$ and $\ket{\Downarrow} \equiv \prod_x \ket{\down}_x$
(which describe SSB of a 0-form $\IZ_2$ symmetry).  
These satisfy condition 2 (since we must flip every spin to get from $\ket{\Uparrow}$ to $\ket{\Downarrow}$, 
but not condition 1 (since \eg~$\bra{\Uparrow} Z_1 \ket{\Uparrow} = 1 = - \bra{\Downarrow} Z_1 \ket{\Downarrow}$).  
On the other hand, the fact that condition 2 is satisfied is a basis-dependent statement -- it is not true for a linear combination of these states: 
$ Z_1 \( \ket{\Uparrow} + \ket{\Downarrow} \) = - \( \ket{\Uparrow} - \ket{\Downarrow}\) $.  

In contrast, for SSB of 1-form symmetry we'll see that we have condition 1 (in some basis) but not necessarily condition 2.
Note that if condition 1 holds in every basis of the groundstate subspace, then condition 2 holds in every basis, and vice versa.  
Moreover, if both 1 and 2 hold in some basis, then they both hold in any basis.  

Proposition: If $\ket{\gs_1} = W_C \ket{\gs_2}$ for a one-form symmetry operator $W_C$, then condition 1 holds.  

Proof: For all local operators $\CO$ 
\bea 
\bra{\gs_2} \CO \ket{gs_2} &=&  \bra{\gs_1} W_C^\dagger \CO W_C^\apd 
\ket{\gs_1} 
\\ =
\bra{\gs_1} W_{C'}^\dagger \CO W_{C'}^\apd 
\ket{\gs_1} 
& =&  \nonumber
\bra{\gs_1} W_{C'}^\dagger  W_{C'}^\apd
  \CO\ket{\gs_1} 
\\ & =&  
\bra{\gs_1} \CO\ket{gs_1} . 
\nonumber\eea
In the second step we used the topological property of $W_C$ to deform it (if necessary) to avoid the support of the local operator $\CO$. \thump

So SSB of 1-form symmetry implies that, in some basis, condition 1 holds.  

Proposition: If in addition, there's a second 1-form symmetry $V_{\hat C}$ that acts as an order parameter to distinguish the 
states $\ket{\gs_1} = W_C \ket{\gs_2}$, 
in the sense that 
\be V_{\hat C} \ket{\gs_a} = e^{ \ii \theta_a} \ket{\gs_a } ~,~~ a=1,2\ee
with $e^{ \ii \theta_1} \neq e^{ \ii \theta_2}$,
then condition 2 also holds.  

Another way to state the hypothesis is: some operator charged under the first one-form symmetry is also topological.  

Proof: 
The idea is similar to the relation between SSB and long-range order:
For all local operators $\CO$,
\bea \bra{\gs_2} \CO \ket{\gs_1} 
&=& \bra{\gs_2} V_{\hat C}^\dagger \CO V_{\hat C}^\apd \ket{\gs_1} e^{ \ii ( \theta_2 - \theta_1 )} 
\nonumber
\\ = 
\bra{\gs_2} V_{\hat C}^\dagger  V_{\hat C}^\apd \CO \ket{\gs_1} e^{ \ii ( \theta_2 - \theta_1 )}  
&= & 
\bra{\gs_2}  \CO \ket{\gs_1} e^{ \ii ( \theta_2 - \theta_1 )}
. \eea
Since $ e^{ \ii (\theta_2 - \theta_1 ) } \neq 1$ by assumption, we conclude that 
$ \bra{\gs_2}  \CO \ket{\gs_1}  $ = 0 .  \thump

A state very much like the Chamon-Castelnovo state is realized upon subjecting the toric code to enough decoherence \cite{Dennis:2001nw,Fan:2023rvp,Bao:2023zry,chen2023separability,Lee:2024yed,Chen:2024knt}.  
For weak-enough decoherence, the topological order survives  \cite{Dennis:2001nw}, a good sign for its usefulness for real quantum information processing.
At a certain threshold of decoherence strength, there is a phase transition to a phase where the quantum information stored in the groundstates is lost.  Strangely, expectation values of observables are completely unchanged (since the decoherence can be described by a finite-depth quantum channel), so for example, the Wilson loop still has a perimeter law, and there is still SSB of a one-form symmetry.  
(In fact, the only way to tell that something happened is by looking at quantities nonlinear in the density matrix.)
But beyond the threshold, the {\it other} one-form symmetry, with which the spontaneously-broken one-form symmetry has a mixed anomaly, does not emerge.  So in this phase, there is no algebraic guarantee of a topological degeneracy, and indeed one can check (by various measures nonlinear in the density matrix) \cite{Fan:2023rvp,Bao:2023zry,chen2023separability,Lee:2024yed,Chen:2024knt} that the quantum information is gone.

\bibliographystyle{ucsd}

\bibliography{all.bib}
\end{document}